\documentclass[]{aa}

\usepackage{psfig}
\usepackage{graphicx}
\usepackage{natbib}
\bibpunct{(}{)}{;}{a}{}{,}

\begin{document}

\title{
Dust depletion and abundance pattern in damped Ly$\alpha$ systems: a sample
of Mn and Ti abundances at $z<2.2$\thanks{Based on observations carried out at
the European Southern Observatory (ESO), La Silla, under prog. ID
No. 56.B-0261 and 59.B-0797, and during Science Verification of the
UVES instrument at the ESO Very Large Telescope, Paranal (Chile). Also based
on Hubble Space Telescope data retrieved from the Space Telescope
European Coordinating Facility archive.}}

\titlerunning{Dust depletion and abundance pattern in DLA systems}

\author{C\'edric Ledoux\inst{1}
       \and
        Jacqueline Bergeron\inst{1}
       \and
	Patrick Petitjean\inst{2,3}
}

\offprints{C. Ledoux}

\institute{
   European Southern Observatory, Karl-Schwarzschild Stra\ss e 2,
   D-85748 Garching bei M\"unchen, Germany\\
   \email{cledoux@eso.org, jbergero@eso.org}
	\and
   Institut d'Astrophysique de Paris, 98bis Boulevard Arago,
   F-75014 Paris, France\\
   \email{petitjean@iap.fr}
	\and
   DAEC, Observatoire de Paris-Meudon,
   F-92195 Meudon Principal Cedex, France
}

\date{Received 3 August 2001 / Accepted 4 February 2002}

\abstract{
We analyse a sample of 24 damped Lyman-$\alpha$ (DLA)/moderate DLA systems at
intermediate redshifts, $0.3<z_{\rm abs}<2.2$, all with measurement of the weak Mn\,{\sc ii}
absorption lines, to investigate which elemental ratios could possibly be
used as tracers of either dust depletion or nucleosynthesis effects.
We applied a component-by-component analysis to the five systems of the sample
with new observations and, using this procedure, re-analyzed data
gathered from the literature whenever possible. We show that the
standard method which uses column densities integrated over the whole
absorption profiles could substantially underestimate the abundance of
rare elements relative to Fe. We find a correlation between the
observed [Si/Fe] and [Zn/Fe] ratios, present in our sample at the $2.9\sigma$
significance level. This correlation is fully consistent with a dust
depletion sequence only for a Galactic warm disk cloud or halo cloud
depletion pattern. The correlation between [Mn/Fe] and [Zn/Fe], detected at
the $3.2\sigma$ significance level, cannot be accounted for by any dust
depletion sequence: it implies either variations of the intrinsic Mn
abundance relative to Fe from $-0.3$ to $+0.1$ dex and/or a relation between
depletion level and metallicity. The correlation between [Mn/Fe] and
metallicity ($2.6\sigma$ significance level) strengthens the assumption
of intrinsic variations of [Mn/Fe] although some marginal correlation
between [Zn or Si/Fe] and [Zn/H] is present as well. Extension of the sample
toward low metallicity is needed to confirm the correlation between
depletion level and metallicity. The variations of [Ti/Fe] vs. [Zn/Fe]
cannot be fitted by a single dust depletion sequence either. We then adopt a
warm disk cloud or halo cloud depletion pattern and compare the resulting
dust-corrected abundance ratios to those observed in Galactic and SMC stars.
At high metallicity, [Fe/H$]_{\rm dc}\ga -0.5$, the intrinsic abundance
pattern of Si, Ti, Cr and Mn in DLA absorbers closely follows the trends
observed in Galactic stars and these absorbers should thus have a chemical
evolution similar to that of our Galaxy. At lower metallicity, some absorbers
do follow the trends present in Galactic stars but a substantial fraction
of them have elemental ratios (in particular [Si/Fe$]_{\rm dc}$
and [Mn/Fe$]_{\rm dc}$) closer to the solar values than Galactic stars.
This could be explained by a larger contribution of type Ia supernovae to the
chemical enrichment of these DLA absorbers than in Galactic stars of similar
metallicity. This metal-poor DLA absorber population could
trace H\,{\sc i}-rich dwarf galaxies.
}

\maketitle

\section{Introduction}\label{intro}

The association of damped Lyman-$\alpha$ (DLA) systems with galaxies, as first
suggested by \citet{1986ApJS...61..249W}, has been demonstrated at low and
intermediate redshifts by the identification of a small sample of DLA systems
at $z_{\rm abs}\la 1$ \citep{1997A&A...321..733L}. These absorbers have been
extensively used to probe the chemical history and the properties of the
gaseous halos of distant galaxies. Their metallicity was first derived
from high-resolution spectroscopic observations of Zn\,{\sc ii}
absorption lines as this element is not significantly depleted onto
dust grains in the Galactic interstellar medium (ISM). The derived
$N($H\,{\sc i}$)$-weighted mean metallicity of DLA systems at
$z_{\rm abs}\ga 2.5$ is about $1/13Z_\odot$ with a large spread of two orders
of magnitude \citep{1997ApJ...486..665P}. The value found at
$z_{\rm abs}< 1.5$ is only slightly larger, $\sim 1/10Z_\odot$
\citep{1999ApJ...510..576P}. These low metallicities point toward young
galaxies, although the exact nature of high-redshift DLA systems is still
controversial \citep[see][]{1997ApJ...487...73P,1998A&A...337...51L}. The wide
range in metal enrichment and lack of significant evolution could be due
to the presence of various populations of galaxies of different
morphological types among the DLA absorbers, thus different
star-formation histories.

The underabundance of Cr relative to Zn has been first attributed to
the selective depletion of Cr onto dust grains \citep{1994ApJ...426...79P}
by analogy to the Milky Way. Indeed, the high elemental abundance ratios
[Zn/Cr] ($\equiv\log [N($Zn$)/N($H$)]-\log [N($Cr$)/N($H$)]$) and/or [S/Cr]
in DLA systems with detected H$_2$ molecules demonstrates the presence of
dust in these DLA
absorbers \citep{2002A&Asubmitted..L,2002MNRASsubmitted..P}. However,
this interpretation has been challenged by \citet{1996ApJS..107..475L}
and \citet{1997ApJ...474..140P}. These authors studied the abundances of a
large number of elements and suggested that the observed relative
abundances may reflect the nucleosynthesis pattern of type II supernovae
enrichment. It seems that several groups now tend to agree that both
effects are present and that the [Cr/Zn] ratio is primarily an indicator of
dust depletion
\citep{1997ApJ...478..536P,1997ApJ...484L...7K,1998ApJ...493..583V,1999ApJS..121..369P}.

The main goal of this paper is to analyze a large sample of DLA systems at
intermediate redshift to investigate these issues further. We search for
a clear depletion pattern, in particular for elements less depleted than Fe
in the Galactic ISM other than Zn, as Si. We then concentrate our analysis
on Mn and Ti, elements which have not been extensively studied so far. Our
sample of 24 DLA/moderate-DLA absorbers combines new high spectral resolution
data on five systems at $z_{\rm abs}<1.2$ with previously published data
on DLA/moderate-DLA systems at $0.3<z_{\rm abs}<2.2$. In all cases, the
wavelength range of the expected, associated Mn\,{\sc ii} transition lines
has been observed.

We describe our observations in Sect.~\ref{obser} and present our analysis of
individual absorbers and our measurements in Sects.~\ref{ionic}
and \ref{elabs}, emphasizing the importance of
component-by-component analysis. In Sect.~\ref{obpat}, we discuss the
abundance pattern observed in DLA systems. As explained
in Sect.~\ref{deple}, we then correct the observed abundance ratios for dust
depletion effects to finally draw our conclusions in Sect.~\ref{nucle}.

\section{Observations and data reduction}\label{obser}

\subsection{HST FOS spectra}\label{fos}

UV spectra were retrieved from the HST Faint Object Spectrograph (FOS) archive
to determine the H\,{\sc i} column density from the Lyman-$\alpha$ line of the
metal-rich absorbers toward Q\,0058$+$019 (PHL\,938), Q\,0453$-$423
and Q\,(PKS)\,2128$-$123. The observation log is summarized in
Table~\ref{obsetab}. The spectra were reduced again using the best
reference files available for on-the-fly re-calibration as provided by
the Hubble Space Telescope European Coordinating Facility data
reduction pipeline.

In the cases of Q\,0058$+$019 and Q\,2128$-$123, it was necessary to correct
for scattered light for bringing the cores of the Lyman-$\alpha$ lines to the
zero flux level (correction $<$10\%).
%The corresponding part of the spectra normalized to the quasar continua are
%shown in Fig.~\ref{q0058lyap}.
The signal-to-noise ratio in the adjacent continuum is of the order of 5 and
10 for Q\,0058$+$019 and Q\,2128$-$123 respectively. For Q\,0453$-$423, no
flux is actually observed in the G190H spectrum as is also the case shortward
of about 3020 \AA\ in the G270H spectrum, thereby revealing the existence
of an optically thick Lyman-limit system at $z_{\rm abs}\approx 2.3$ along
the line of sight.

\subsection{VLT UVES data}\label{uves}

High-resolution, high signal-to-noise ratio spectra were obtained with the
UV and Visible Echelle Spectrograph (UVES) at the Kueyen VLT-UT2 8.2
m telescope on Cerro Paranal Observatory during UVES Science Verification
(SV; February 6-17, 2000). Combining two standard dichroic settings: Dic\#1
(grisms B346nm, R580nm) and Dic\#2 (grisms B437nm, R860nm), to observe with
both spectrograph arms, the full optical range was covered from $\sim 3050$ to
10000 \AA\ with only a small gap between the two red CCDs
(see Table~\ref{obsetab}). A slit width of $0\farcs 9$ (resp. $0\farcs 8$)
in the Blue (resp. Red) was used in good seeing conditions and of $1\farcs 0$
when the seeing was poor, yielding a spectral resolution $42500<R<50000$.

The calibrated data for Q\,1122$-$168 was made available to the ESO
community in May 2000. The data reduction was performed by the SV team using
the UVES pipeline \citep{2000Msngr.110...31B}, available as a context of
the MIDAS software. The main characteristics of the pipeline is to perform
a precise inter-order background subtraction, especially for
master flat-fields, and to allow for an optimal extraction of the
object signal rejecting cosmic rays and performing sky-subtraction at the same
time. The pipeline results were checked step by step. We then converted
the wavelengths of the reduced spectra to vacuum-heliocentric values
and scaled, weighted and combined individual 1D exposures using the
NOAO {\it onedspec} package of the IRAF software. The resulting unsmoothed
spectra were re-binned to the same wavelength step (0.0415
\AA\ pix$^{-1}$), yielding a signal-to-noise ratio per resolution element
of 100 (resp. 37) at 5500 \AA\ (resp. 4000 \AA ).

\subsection{ESO CASPEC data}\label{caspec}

High-resolution spectroscopy data of three
quasars, Q\,0058$+$019, Q\,0453$-$423 and Q\,2128$-$123, were obtained
with the CASPEC echelle spectrograph at the ESO 3.6 m telescope at La Silla
Observatory on November 18-21, 1995 and September 23-26, 1997. The
observational details are summarized in Table~\ref{obsetab}. A spectral range
of 1200 \AA\ was covered per setting. The instrumental resolution FWHM is
0.16 \AA\ (or 11 km s$^{-1}$) and the accuracy in the wavelength calibration
is of the order of one tenth of the resolution.

The data were reduced and analyzed using the {\it echelle} package of MIDAS.
The object-flux weighted median of individual frames and cosmic-ray removal
were performed simultaneously. The sky spectrum was difficult to extract
in the blue due to the small spacing between orders. We thus carefully fitted
the zero level to the bottom of the saturated lines in the extracted 1D
spectra. The uncertainty in the flux zero level is about 5\%. The mean
signal-to-noise ratio of the final spectra
%, computed as a function of
%wavelength from photon statistics after subtraction of the spectral lines,
is typically 10, 17 and 28 down to about 4000 \AA , for Q\,0058$+$019,
Q\,0453$-$423 and Q\,2128$-$123 respectively.

\section{Ionic column densities}\label{ionic}

Ionic column densities were derived from a least-squares technique using the
{\it fitlyman} program \citep{1995Msngr..80...37F} in which Voigt profiles
are convolved with the instrumental point spread function.
%For the high-resolution spectra, the identified absorption features have
%a $5\times$FWHM$\times\sigma$ equivalent width limit ($\sigma$ is the
%noise rms in the adjacent continuum).
The metal line profiles 
%associated with the previously fitted Lyman-$\alpha$ lines
%and at $\lambda _r>1800$ \AA\
were fitted consistently assuming the same velocity for all the ions of
a given component and a pure turbulent broadening. Whenever available, we used
updated oscillator strength values \citep[see][]{1996ARA&A..34..279S}, or
else the compilation by \citet{1991ApJS...77..119M}, except for
the Fe\,{\sc ii}\,$\lambda\lambda$2344,2374 lines for which we used
the laboratory measurements of \citet{1996ApJ...464.1044B}. The column
densities of Mg\,{\sc i}, Mg\,{\sc ii}, Si\,{\sc ii}, Ca\,{\sc ii},
Ti\,{\sc ii}, Cr\,{\sc ii}, Mn\,{\sc ii}, Fe\,{\sc ii} and Zn\,{\sc ii} were
derived simultaneously and the results are shown in Table~\ref{paratab}.
In this Table, the standard deviations given in parentheses for
unsaturated line components were also obtained with {\it fitlyman} from a
simultaneous fit of all the lines observed in a given absorber; they should be
realistic if the above assumed model is correct. In what follows, we adopt the
Solar abundances given by \citet{1996ARA&A..34..279S}.

We now comment on the five individual systems of the sample with
new observations.

\subsection{Q\,0058$+$019, $z_{\rm abs}=0.6125$}

The total neutral hydrogen column density of this absorber was derived from
the available HST FOS spectrum.
%taking carefully into account the complex emission continuum of the QSO near the damped Lyman-$\alpha$ line.
We get $\log N($H\,{\sc i}$)=20.14\pm 0.09$.
%which is consistent with the previously quoted value.
We adopted the column density estimated by \citep{2000ApJ...532...65P}
for Fe\,{\sc ii}
%(as well as for Zn\,{\sc ii} and Cr\,{\sc ii})
as it is based on the high signal-to-noise ratio detection of the weak,
unsaturated Fe\,{\sc ii}\,$\lambda\lambda$2249,2260 lines in a Keck HIRES
spectrum. Our Mn\,{\sc ii} measurement yields [Mn/Fe$]=-0.15\pm 0.06$, which
is slightly sub-solar. A $5\sigma$ detection of Ti\,{\sc ii}\,$\lambda$3384
was reported by \citet{2000ApJS..130...91C} using new Keck HIRES data.
The measured column density,
$\log N($Ti\,{\sc ii}$)=12.53\pm 0.08$ (Churchill, private communication),
leads to an elemental ratio [Ti/Fe$]=-0.10\pm 0.09$.

\subsection{Q\,0453$-$423, $z_{\rm abs}=0.7260$ and $z_{\rm abs}=1.1496$}

As mentioned in Subsect.~\ref{fos}, the existence of an optically thick Lyman
limit system at $z_{\rm abs}\approx 2.3$ prevents a direct determination
of the total H\,{\sc i} column density of the $z_{\rm abs}=0.7260$ and 1.1496
strong metal-line systems. However, in both systems,
$w_{\rm r}($Fe\,{\sc ii}\,$\lambda$2600$)/w_{\rm r}($Mg\,{\sc ii}\,$\lambda$2796$)$ is of order unity which, together with the strength of
the Mg\,{\sc ii} lines, $w_{\rm r}($Mg\,{\sc ii}\,$\lambda$2796$)>1$ \AA ,
implies that $N($H\,{\sc i}$)$ is well in excess of a few $10^{19}$ cm$^{-2}$.
Both absorbers should thus be moderate-DLA if not DLA
systems \citep[see][]{1986A&A...169....1B,2000ApJS..130....1R}.
For the system at $z_{\rm abs}=1.150$, the Fe\,{\sc ii} line profile
is complex spanning about 550 km s$^{-1}$. A simultaneous least-square fit of
the Fe\,{\sc ii}\,$\lambda\lambda$2249,2260 and Fe\,{\sc ii}\,$\lambda$2344
profiles (the latter constraining the line parameters of the components 5 to
8; see Table~\ref{paratab}) yields a {\it total} Fe\,{\sc ii} column
density, $\log N($Fe\,{\sc ii}$)=15.18\pm 0.08$, thus probably a high
metallicity ([Fe/H$]>0.5$).
%in agreement with that
%determined by \citet{1994A&A...283..759P} from the observations of
%the Fe\,{\sc ii}\,$\lambda\lambda\lambda$2382,2586,2600 lines.
For the system at $z_{\rm abs}=0.726$, we
measured $\log N($Fe\,{\sc ii}$)=14.54\pm 0.04$. The Mn abundances relative
to Fe derived using the method outlined in Sect.~\ref{elabs}
are over-solar, [Mn/Fe$]=+0.12\pm 0.09$ and $+0.34\pm 0.20$
at $z_{\rm abs}=0.726$ and 1.150 respectively (see Table~\ref{ablitab}).

\subsection{Q\,1122$-$168, $z_{\rm abs}=0.6819$}\label{q1122}

\begin{figure*}
\hbox{
\psfig{figure=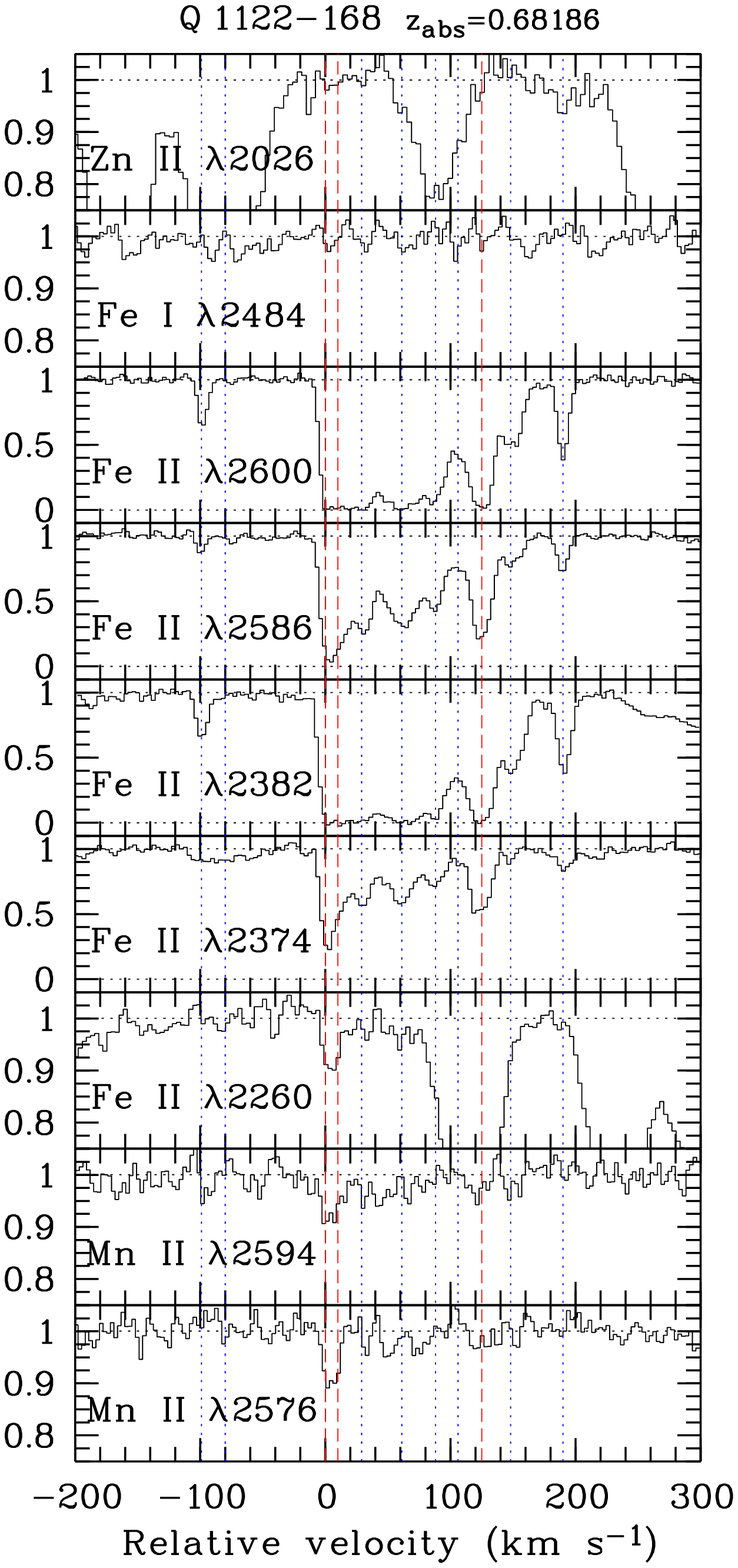,width=8.7cm,clip=,bbllx=47.pt,bblly=32.pt,bburx=409.pt,bbury=798.pt,angle=0.}\hspace{0.20cm}
\psfig{figure=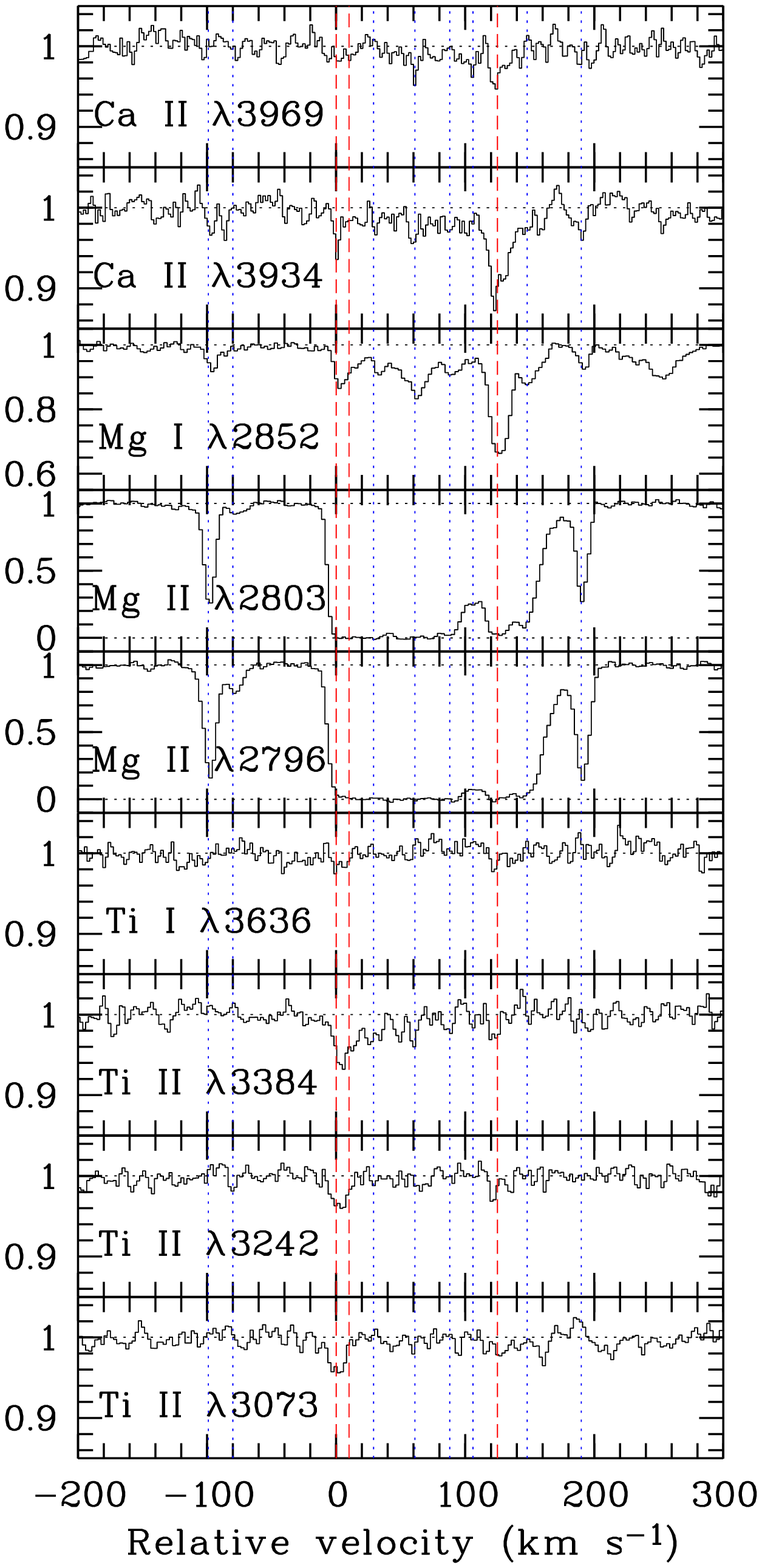,width=8.7cm,clip=,bbllx=47.pt,bblly=32.pt,bburx=409.pt,bbury=798.pt,angle=0.}}
\caption[]{Low-ionization line profiles from the $z_{\rm abs}=0.68186$ DLA
system in the normalized VLT-UVES spectra of Q\,1122$-$168.
Individual components (see Table~\ref{paratab}) are marked by vertical lines
and the dashed lines indicate the main components used to derive elemental
ratios (see text and Table~\ref{ablitab}). For this absorber, we
have considered two subsystems.}
\label{q1122plot}
\end{figure*}

This DLA system was recently studied by \citet{2000A&A...363...69D} using
ESO 3.6 m CASPEC and Keck HIRES data. From an HST FOS spectrum, these
authors derived a total H\,{\sc i} column density
$\log N($H\,{\sc i}$)=20.45\pm 0.05$ from a fit of the Lyman-$\alpha$
and Lyman-$\beta$ lines. In our very high signal-to-noise ratio UVES spectra,
we measured accurate column densities for Fe\,{\sc ii}, Mn\,{\sc ii}
and Ti\,{\sc ii}, and placed a stringent upper limit on the column density of
Zn\,{\sc ii} (see Table~\ref{paratab}).
%The Mg\,{\sc ii} doublet was
%not fitted due to the saturation of the profiles except for the
%satellite component at $V\sim -100$ km s$^{-1}$.
The Ca\,{\sc ii} lines are optically thin and their profile closely
follows that of the Mg\,{\sc i}\,$\lambda$2852 line from $-100$ to $+200$ km
s$^{-1}$ (see Fig.~\ref{q1122plot}).
%These two elements indeed trace dense and neutral gas.
The Ca\,{\sc ii} over Mg\,{\sc i} column density ratio is indeed
essentially the same for all the components in which both species are
detected, $\log (N($Ca\,{\sc ii}$)/N($Mg\,{\sc i}$))\simeq -0.28$, and
is consistent with the prediction for warm gas with low
depletion \citep[e.g.][]{1997A&A...327...47V}.

The Fe abundance derived from the weaker transitions of Fe\,{\sc ii} is
[Fe/H$]=-1.40\pm 0.05$. The non-detection of Zn\,{\sc ii}
yields [Zn/Fe$]<+0.29$ and $<+0.39$ for components $3+4$ and 9
respectively (see Table~\ref{paratab} and \ref{ablitab}). This DLA system thus
appears to be essentially dust-free \citep[see also][]{2000A&A...363...69D}.
The Mn\,{\sc ii} column densities we derived for the components $3+4$ and 9
lead to sub-solar elemental ratios, [Mn/Fe$]=-0.25\pm 0.09$ and
[Mn/Fe$]\le -0.40\pm 0.12$ respectively. Note that these values estimated for
individual components are 0.3 to 0.4 dex higher than those integrated over
the whole profile as given by \citet{2000A&A...363...69D}. This is due to
a non-negligible contribution to the total Fe\,{\sc ii} column density
of components detected in Fe\,{\sc ii} but not in Mn\,{\sc ii}. {\it This
clearly demonstrates that systematic of this kind can dominate the
abundance ratio variations in DLA systems}. The Ti abundance is derived from
a simultaneous fit of
the Ti\,{\sc ii}\,$\lambda\lambda\lambda$3073,3242,3384 lines. We get
[Ti/Fe$]=+0.06\pm 0.09$ and [Ti/Fe$]\le -0.04\pm 0.12$ for
components $3+4$ and 9 respectively.
%Contrary to [Mn/Fe], these values
%are consistent with solar abundance ratios.

\subsection{Q\,2128$-$123, $z_{\rm abs}=0.4298$}

Our determination of the total H\,{\sc i} column density from the HST
FOS spectrum is $\log N($H\,{\sc i}$)=19.37\pm 0.08$. Although the
Lyman-$\alpha$ line has damped wings, this value is lower than the
threshold usually adopted for DLA systems ($\ga 20.3$). Since
$\log N($H\,{\sc i}$)\simeq 19.4$ corresponds to an optical depth at the
Lyman limit $\tau_{\rm LL}\simeq 150$, ionization corrections should be small
\citep[e.g.][]{1995MNRAS.276..268V}. Furthermore, we concentrate
our discussion below on relative abundances and these elements have
negligible differential ionization corrections
\citep[see][]{2001ApJ...557.1007V}.

In our ESO 3.6 m CASPEC spectrum, the line profiles
from low-ionization species show two separated components
spanning $\Delta V\simeq 75$ km s$^{-1}$, the strongest component
being located on the profile red side.
%(see Fig.~\ref{q2128plot}).
%Mg\,{\sc i}, Mg\,{\sc ii} and Fe\,{\sc ii}
%Two separated, main components are clearly present which excludes the single
%velocity-component model of \citet{1992ApJ...391...48L}.
%The published results for the Fe\,{\sc ii}\,$\lambda\lambda$2586,2600 doublet
%\citep{1990A&A...231..309P} and the Fe\,{\sc ii}\,$\lambda$1608
%line \citep{1998ApJS..118....1J}
%%$w_{\rm obs}($Fe\,{\sc ii}\,$\lambda$1608$)=0.14\pm 0.03$ \AA\
%were used to derive the Fe\,{\sc ii} column density. From a curve of
%growth analysis, we find $\log N($Fe\,{\sc ii}$)=14.03\pm 0.19$ and
%$b=11.2\pm 0.5$ km s$^{-1}$. This result is confirmed by our
A two velocity-component model was used to fit the Fe\,{\sc ii} and
Mg\,{\sc ii} doublets and the Mg\,{\sc i} line detected in the observed
wavelength range, 3600-4100 \AA\ (see Table~\ref{paratab}). The velocity
broadening of the main component is constrained by
the Fe\,{\sc ii}\,$\lambda$2586 profile, whereas that of the
satellite component is determined from the strong Mg\,{\sc ii} doublet.
Since the latter Fe\,{\sc ii} line may be saturated, we give only a lower
limit on the column density: $\log N($Fe\,{\sc ii}$)>14.08$ in the
main component. From the non-detection of the Mn\,{\sc ii} lines, we
get [Mn/Fe$]<-0.03$ (see Table~\ref{ablitab}). From
$\log N($Ti\,{\sc ii}$)<11.21$ (Churchill, private communication), we derive
[Ti/Fe$]<-0.29$.
%and since Ti, as Mn, has little or no ionization correction relative to Fe,
%This is definitively sub-solar.

\section{The DLA absorber sample}\label{elabs}

In order to investigate the abundance pattern of DLA systems and study
dust depletion effects, we have built a sample of 24 DLA/moderate DLA (i.e.
$\log N($H\,{\sc i}$)>20.3$/$19.4\leq\log N($H\,{\sc i}$)<20.3$) systems
at $0.3<z_{\rm abs}<2.2$, all with measurements of the Mn\,{\sc ii}
absorption lines. The sample, as shown in Table~\ref{ablitab}, includes
seven DLA candidates selected on the basis of their large
$0.6\la w_r($Fe\,{\sc ii}\,$\lambda$2600$)/w_r($Mg\,{\sc ii}\,$\lambda$2796$)\la 1.0$
equivalent width ratio. We give the measured abundances of Si, Ti, Cr, Mn and
Zn relative to Fe. The previously analyzed Mn samples are those of
\citet{1996ApJS..107..475L}, which includes seven Mn\,{\sc ii} systems
at $0.86\leq z_{\rm abs}\leq 2.14$ of which three are associated with DLA
candidates, and \citet{2000ApJ...532...65P}, which includes
six Mn\,{\sc ii} systems at $0.61\leq z_{\rm abs}\leq 1.42$ of which one is in
common with those of Lu et al. In our sample, there are nine Ti\,{\sc ii}
detections and three upper limits for systems at
$0.43\leq z_{\rm abs}\leq 1.96$. The Ti abundance measurements in the
DLA systems toward Q\,0058$+$019, Q\,0454$+$039, Q\,1622$+$238 and
Q\,2128$-$123 were kindly made available to us by C. W. Churchill. They were
derived from Keck HIRES spectra obtained for another
project \citep{2000ApJS..130...91C,2000ApJ...543..577C}. The sample of
six Ti\,{\sc ii} absorbers detected by \citet{2001ApJS..137...21P} covers a
somewhat higher redshift range, $1.78\leq z_{\rm abs}\leq 2.10$.

The relative abundances in the gaseous phase of DLA systems are usually
determined from column densities integrated over the entire
absorption profiles. This procedure leads to unbiased results only when
all the components observed in the profiles of strong absorption lines are
also detected in weaker transition lines. However, the optical depth of most
of the Fe\,{\sc ii} lines are larger than those of
the Mn\,{\sc ii}, Ti\,{\sc ii}, Cr\,{\sc ii} and Zn\,{\sc ii} transitions.
As the overall profiles of the weaker transitions are usually less extended
than those of the Fe\,{\sc ii} lines, the above procedure tends to
overestimate the column density of Fe\,{\sc ii} as compared to other species.
For this reason, we have used a different approach. After performing the
Voigt profile fits of all of the absorption profiles, we summed up only the
column densities of the components detected in at least one of the
Mn\,{\sc ii}, Ti\,{\sc ii}, Cr\,{\sc ii} or Zn\,{\sc ii} ions (note that
these ions, when observed, are usually detected at the same time). These
components correspond most often to the strongest part of the profiles.
%in all the absorbers of our sample, except that toward
%Q\,1122$-$168 for which we considered two subsystems.

Our procedure tends to minimize the impact of limitations due to finite
signal-to-noise ratios whereas the standard procedure can lead
to underestimate the abundances of rare elements relative to Fe by $-0.1$
up to $-0.3$ dex (as is the case for the DLA system toward Q\,1122$-$168). For
DLA systems at intermediate redshifts, we also applied this procedure to all
the data taken from the literature when a detailed modeling was available
(see Table~\ref{ablitab}). However, for the estimate of the absolute [Fe/H]
abundances, we include all the components of Fe\,{\sc ii} for comparison with
previous studies.

\section{Observed abundance pattern}\label{obpat}

We present in Table~\ref{ablitab} the abundance of Fe and elemental abundance
ratios relative to Fe for our sample. From the analysis of a DLA sample
at intermediate redshift, \citet{1999ApJ...510..576P} showed that there is no
significant cosmic evolution of DLA metallicity ([Zn/H]) in our
redshift range, $0.3<z_{\rm abs}<2.2$. The elemental abundance ratios
presented in Fig.~\ref{abobs} and Table~\ref{arcetab} do not exhibit
any significant evolution with redshift either. The scatter of these
abundance ratios is very small for [Cr/Fe] and about 1 dex
for [Mn/Fe] and [Zn/Fe].

We examine below the observed abundance patterns to identify the
elements which would clearly help separating the effects
of star-formation history and depletion of refractory elements onto
dust grains.

\subsection{Zn and Cr}

\begin{figure*}
\hbox{
\hspace{3.0cm}\psfig{figure=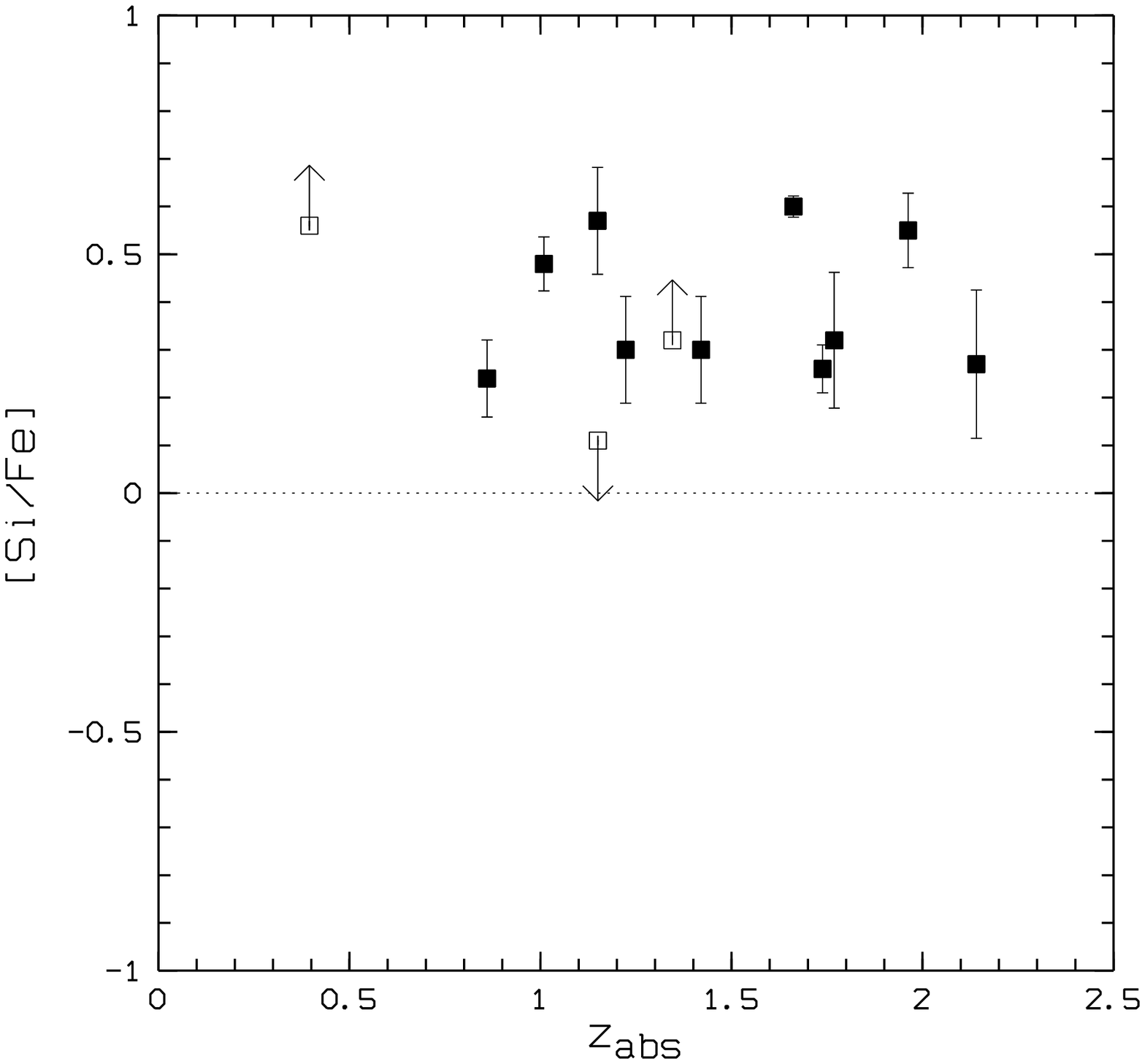,width=5.8cm,clip=,bbllx=40.pt,bblly=295.pt,bburx=555.pt,bbury=780.pt,angle=0.}\hspace{0.2cm}\psfig{figure=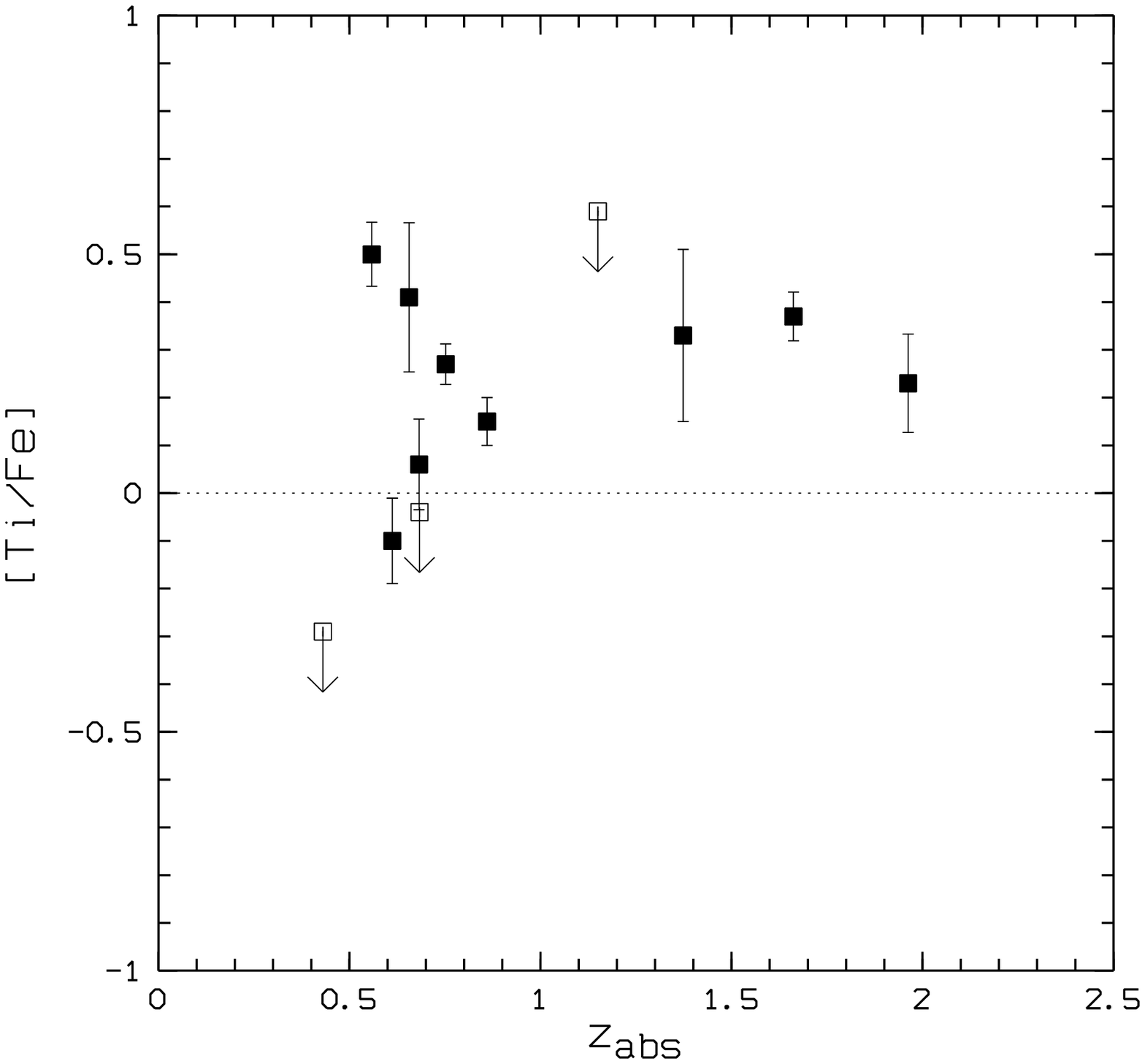,width=5.8cm,clip=,bbllx=40.pt,bblly=295.pt,bburx=555.pt,bbury=780.pt,angle=0.}}
\vspace{0.15cm}
\hbox{
\psfig{figure=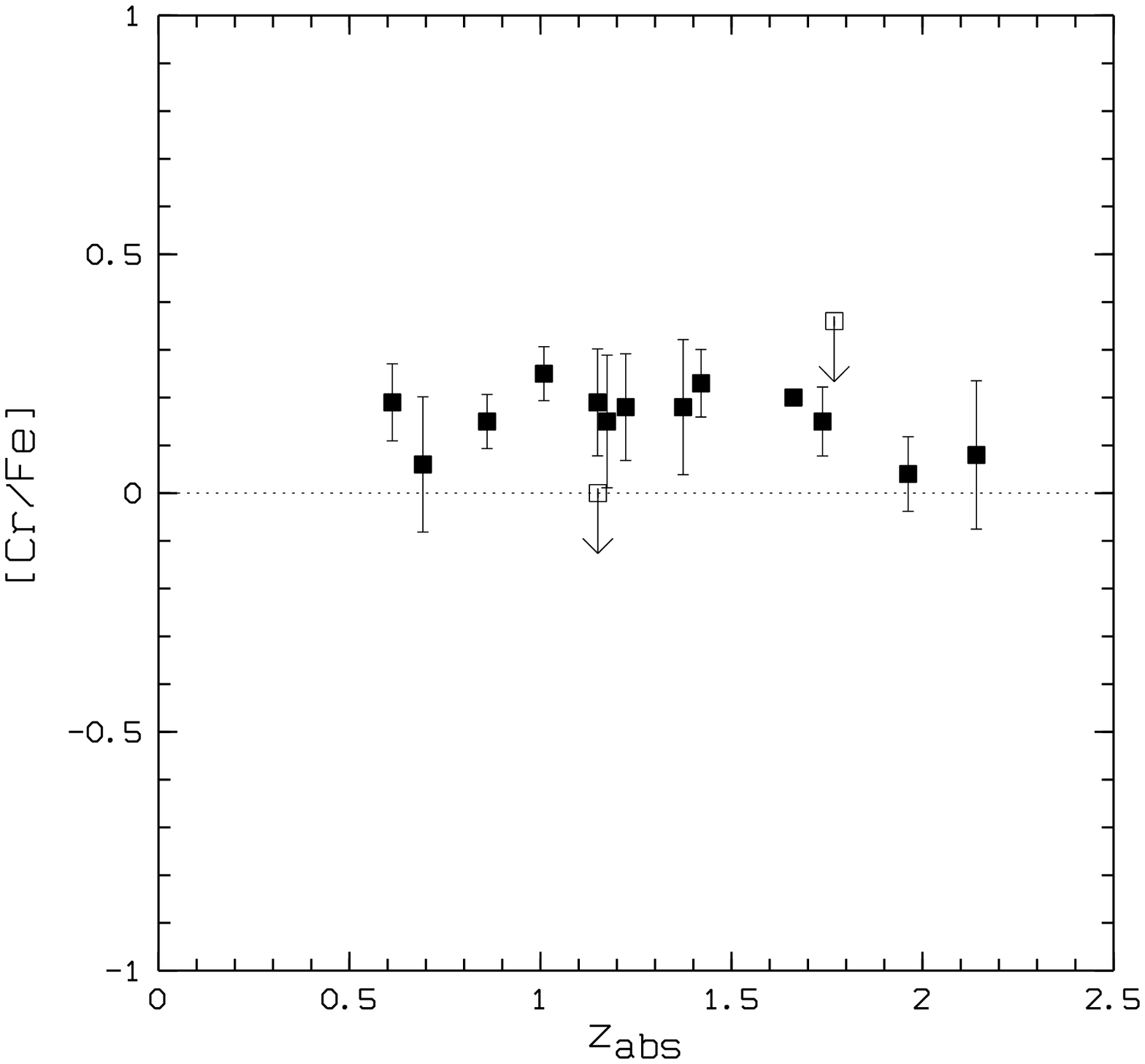,width=5.8cm,clip=,bbllx=40.pt,bblly=295.pt,bburx=555.pt,bbury=780.pt,angle=0.}\hspace{0.2cm}\psfig{figure=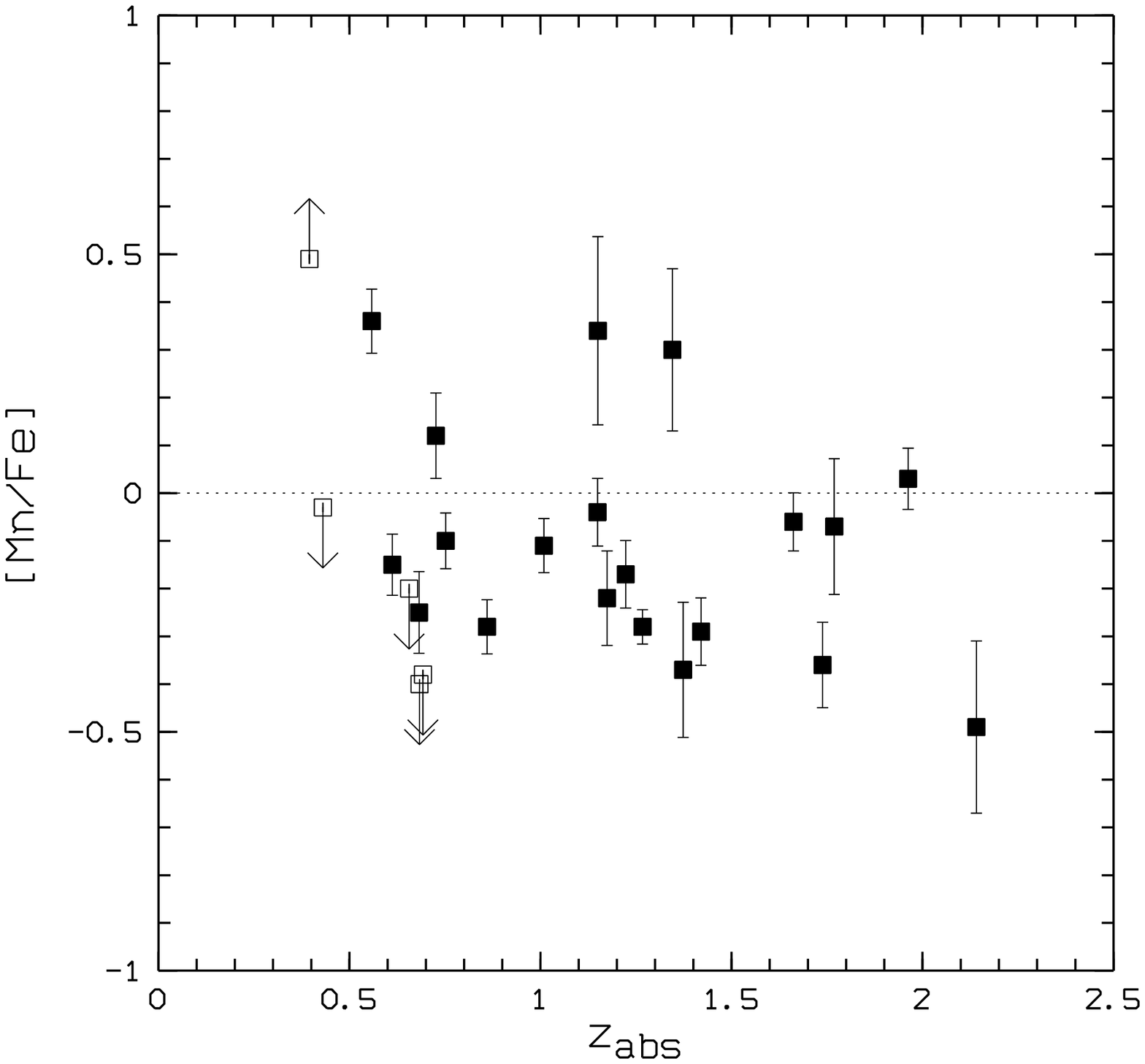,width=5.8cm,clip=,bbllx=40.pt,bblly=295.pt,bburx=555.pt,bbury=780.pt,angle=0.}\hspace{0.2cm}\psfig{figure=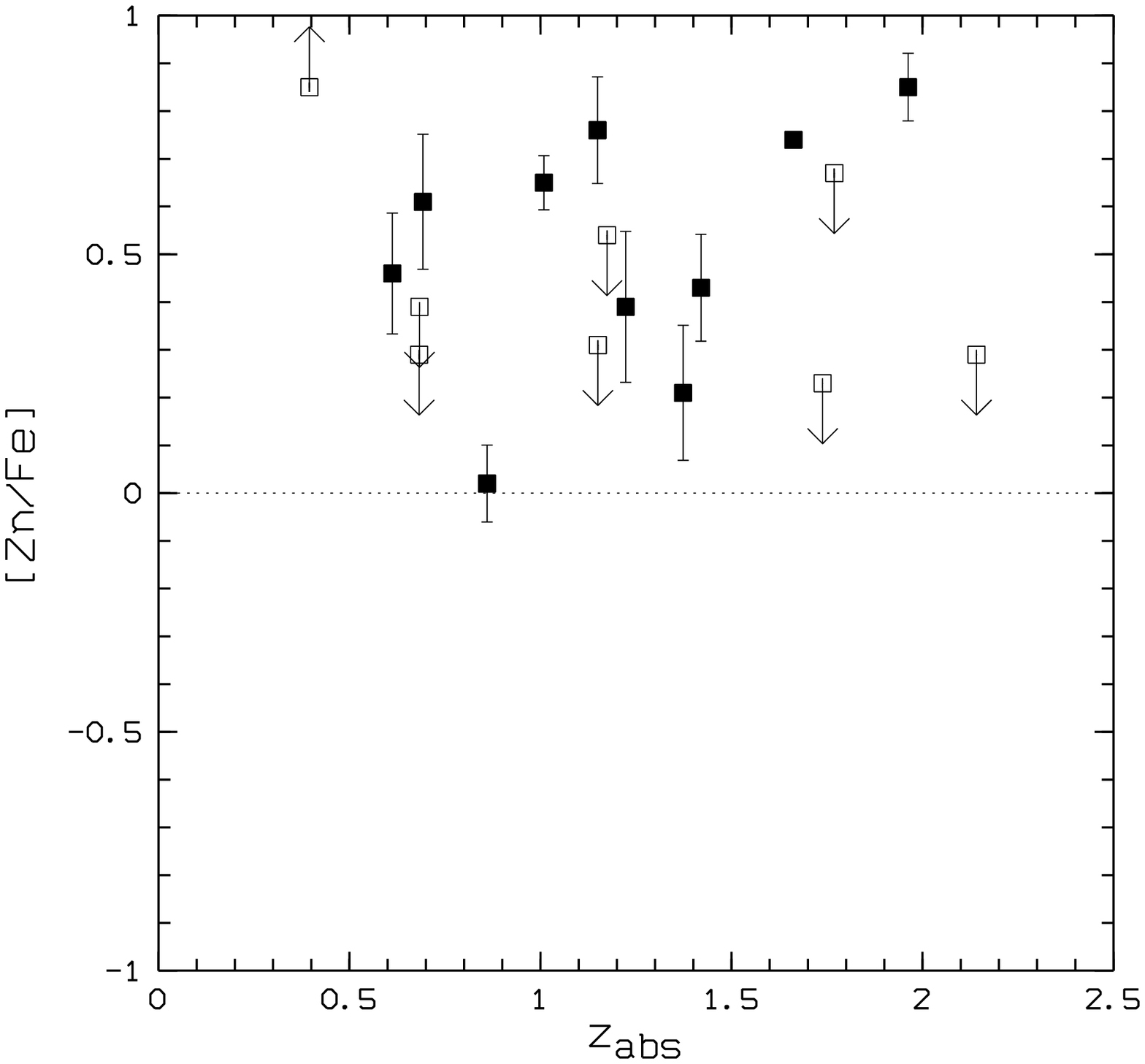,width=5.8cm,clip=,bbllx=40.pt,bblly=295.pt,bburx=555.pt,bbury=780.pt,angle=0.}}
\caption[]{Observed abundance ratios of Si, Ti, Cr, Mn and Zn relative to
Fe versus redshift for DLA systems at $0.3<z_{\rm abs}<2.2$. There is
no apparent cosmic evolution of the abundance ratios in this redshift range.}
\label{abobs}
\end{figure*}

In DLA systems at $0.3<z_{\rm abs}<2.2$, Zn and Cr are both observed to be
overabundant relative to Fe (see Fig.~\ref{abobs}). The spread in [Zn/Fe]
ranges from a solar ratio up to overabundances of about $+0.8$, with a mean
of about $+0.5$ (see Table~\ref{arcetab}). It is unclear whether
an overabundance of Zn relative to Fe can be explained by stellar
nucleosynthesis models. The channels of Zn production in massive stars are not
yet fully understood and the predictions of standard nucleosynthesis models
lie at least a factor of 2 below the observed stellar abundances of Zn
(\citealt{1995ApJS...98..617T,1996ApJ...460..408T}; however see
\citealt{2002ApJ...565..385U}). Additional processes might produce
important amounts of Zn, such as axi-symmetrically deformed explosions in type
II supernovae \citep{1999ApJ...511..341N} and {\it p}-processing in
the neutrino-driven wind following the supernovae explosions
\citep{1996ApJ...460..478H}, but quantitative results depend on uncertain
parameters. Moreover, in Galactic thin disk stars the observed abundance of Zn
follows that of Fe with [Zn/Fe$]=+0.04\pm 0.10$ and in Galactic thick disk and
halo stars the observed [Zn/Fe] ratio is in the
range (0,$+0.2$) \citep[see][ and references
therein]{2000AJ....120.2513P}. Finally, in the Galactic ISM, Zn is
mostly undepleted even in dust-rich regions.

We thus infer that, except maybe for elemental
ratios [Zn/Fe$]_{\rm obs}\la +0.2$ (see also Sect.~\ref{deple}), dust
depletion should be the main effect governing the observed [Zn/Fe] abundance
pattern of DLA systems.

In DLA absorbers, the observed [Cr/Fe] ratio shows very little scatter
whatever the metallicity (see Table~\ref{ablitab}), $z_{\rm abs}$
(see Fig.~\ref{abobs}) or [Zn/Fe] (see Fig.~\ref{deseq}). The overabundance of
Cr relative to Fe ($\sim +0.15$) was already noted
\citep{1996ApJS..107..475L,1999ApJS..121..369P}. It cannot be accounted
for solely by dust depletion since it is present even in systems with only
small amounts of dust ($0\la [$Zn/Fe$]<0.3$) or of low metallicity
([Zn/H]$<-1$). Moreover, Cr and Fe are depleted onto dust grains by fairly
similar amounts in the warm ISM of the
Galaxy, [Cr/Fe]$_{\rm ISM}\simeq +0.13$ \citep[see][]{1996ARA&A..34..279S}
and the ISM of the SMC (see Table~\ref{arcetab}). However, this elemental
ratio is larger in DLA absorbers than in Galactic stars by a factor of
about $+0.15$ dex and close to those observed in the stars of the SMC
(see Table~\ref{arcetab}). In current chemical evolution models, the abundance
of Cr closely follows that of Fe for a wide range of
metallicities \citep[e.g.][]{1995ApJS...98..617T,2000A&A...359..191G}.
These models reproduce the trend observed in Galactic stars but not the
SMC value. Therefore, in DLA absorbers dust depletion effects could
be important, as well as nucleosynthesis ones as for SMC stars.

\subsection{Mn, Si and Ti}

In DLA absorbers, the abundance of Mn relative to Fe spans a larger range
than previously found from smaller samples. For a substantial fraction of our
sample, we get $-0.5\la [$Mn/Fe$]\la 0$, but there are also five systems at
$z_{\rm abs}<1.4$ with [Mn/Fe$]>0$ (see Fig.~\ref{abobs}). Of the latter, only
one is a confirmed DLA absorber with $\log N($H\,{\sc i}$)>20.3$. As can
be seen in Table~\ref{arcetab}, the mean abundance of Mn,
an odd-$Z$ Fe-peak element, in DLA systems is close to that observed
in Galactic metal-poor halo stars
\citep{1995AJ....109.2757M,1996ApJ...471.254R} and the cases with
[Mn/Fe$]\sim 0$ are similar to those of SMC stars. The dust depletion level
of Mn is about the same as that of Fe in Galactic warm halo clouds and
slightly lower in Galactic warm disk clouds \citep{1996ARA&A..34..279S}. From
the spread of values of [Mn/Fe] in DLA absorbers together with the small
relative ionization correction factors for these elements \citep[as well as
for Ti: see e.g.][]{2001ApJ...557.1007V}, we thus infer that
both nucleosynthesis and dust depletion effects are needed to account for this
elemental ratio.

The abundances of the two $\alpha$-elements Si and Ti are larger than solar
in DLA absorbers, with a few cases slightly sub-solar for Ti
(see Fig.~\ref{abobs}). These elements are also overabundant in Galactic
halo and SMC stars (see Table~\ref{arcetab}). The overabundance of Si can be a
consequence of nucleosynthesis effects
\citep{1996ApJS..107..475L,2000ApJ...532...65P} since in Galactic chemical
evolution models the abundance of $\alpha$-elements is enhanced as compared to
solar for metallicities $-2\la [$Fe/H$]\la -1$
\citep[e.g.][]{1995ApJS...98..617T,1997ARA&A..35..503W}. However, it is also
consistent with a dust depletion pattern similar to that observed in Galactic
warm disk and halo clouds (see Sect.~\ref{deple}). As Ti is observed to be
more depleted onto dust grains than Fe in Galactic warm halo clouds,
nucleosynthesis effects are most probably required to account for the
observed over-solar [Ti/Fe] ratios of DLA systems, and this, whatever the dust
content. However, dust depletion could be the dominant factor for the few
cases with observed, sub-solar [Ti/Fe] ratio.

\section{Depletion sequences}\label{deple}

\subsection{Method}\label{method}

When a uniform depletion fraction is assumed, the intrinsic abundances
are simply given by:
\begin{equation}\label{pettini}
10^{\left[\frac{\rm X}{\rm Zn}\right]_{\rm dc}} = 10^{\left[\frac{\rm X}{\rm Zn}\right]_{\rm obs}}\cdot 10^{\left[\frac{\rm Zn}{\rm Cr}\right]_{\rm obs}},
\end{equation}
This depletion correction follows the observed composition of the gaseous
phase, thus does not change the relative abundances. It should be used
only for low depletion factors, [Zn/Fe$]_{\rm obs}\la +0.3$, i.e. for only
a small fraction of our sample.

A simple method which takes into account the dust composition was developed
by \citet{1998ApJ...493..583V}. Various depletion patterns can be used with
different fractions, $f_{\rm X}$, of an element X locked onto dust grains.
From Vladilo's Eqs.~(17) and (18), introducing explicitly in Eq.~(18) the
intrinsic [Zn/Fe] elemental ratio to allow for non-solar values, we derive the
following general relation:
\begin{equation}\label{vladilo2}
10^{\left[\frac{\rm X}{\rm Fe}\right]_{\rm obs}} = 
\frac{10^{\left[\frac{\rm X}{\rm Fe}\right]_{\rm dc}}-\frac{\tilde{k}}
{\tilde{Z}}f_{\rm X}\cdot 10^{\left[\frac{\rm Zn}{\rm Fe}\right]_{\rm dc}}
}{1-\frac{\tilde{k}}{\tilde{Z}}f_{\rm Fe}\cdot 
10^{\left[\frac{\rm Zn}{\rm Fe}\right]_{\rm dc}}},
\end{equation}
where the label ``dc'' refers to intrinsic (dust corrected) elemental ratios,
and $\tilde{k}$/$\tilde{Z}$ is the dust-to-metal ratio normalized to the
Galactic value.

To compute a ``depletion sequence'', [X/Fe$]_{\rm obs}$
versus [Zn/Fe$]_{\rm obs}$, we need to assume a value for [Zn/Fe$]_{\rm dc}$.
Since in Galactic stars of low metallicity and in SMC stars this ratio is
observed to be slightly over-solar, $0\la [$Zn/Fe$]\la +0.2$
(see Table~\ref{arcetab}), we discuss in Sect.~\ref{nucle} the effect of a
slight overabundance of Zn relative to Fe on the dust depletion correction.

\subsection{Results}\label{results}

We present in Fig.~\ref{deseq} the observed abundance ratios of Si, Ti, Cr and
Mn relative to Fe versus [Zn/Fe$]_{\rm obs}$ together with depletion sequences
for various depletion patterns, namely those of Galactic cold disk
clouds, Galactic warm disk and Galactic warm halo clouds as provided by
\citet{1999ApJS..124..465W}. The observations show a correlation between
[Si/Fe$]_{\rm obs}$ and [Zn/Fe$]_{\rm obs}$, detected at the $2.9\sigma$
significance level using a Kendall rank correlation test. A similar trend is
present in the higher redshift sample analyzed by \citet{1999ApJS..121..369P}.
Such a correlation is fully consistent with a dust depletion sequence.
%Moreover, as Zn is less depleted than Si in the Galactic ISM,
%[Zn/Fe$]_{\rm obs}$ increases more rapidly than [Si/Fe$]_{\rm obs}$.
The best agreement between model and observations is obtained for the
Galactic warm disk and warm halo cloud patterns. These patterns are similar
to those observed in the ISM of the SMC \citep{2001ApJ...554L..75W}. The
Galactic warm halo cloud pattern is also close to that observed in
the Magellanic Bridge \citep{2001ApJ...551..781L} where the overall
metallicity may be even lower than in the SMC. In addition, the data allows
for intrinsic elemental ratios
slightly over-solar, 0$<[$Si/Fe$]_{\rm dc}<+0.2$. A good depletion sequence
fit is then also obtained for
[Si/Fe$]_{\rm dc}\simeq [$Zn/Fe$]_{\rm dc}=+0.1$. We thus conclude
that, although some nucleosynthesis effects could be present
(see Sect.~\ref{nucle}), the variations of the observed [Si/Fe]
elemental ratio should mainly be due to dust depletion.

Being fairly constant down to low depletion levels, the [Cr/Fe$]_{\rm obs}$
ratio is not well fitted by a depletion sequence, although a large fraction of
the data points are roughly consistent with dust depletion if an intrinsic
overabundance of Cr relative to Fe of $+0.1$ dex is considered (see
Fig.~\ref{deseq}). This small overabundance is not observed in Galactic
metal-poor stars \citep{1995AJ....109.2757M}, whereas it is present in
both the ISM and the stars of the SMC. Although current chemical
evolution models reproduce satisfactorily a constant
[Cr/Fe$]\approx 0$ \citep[see
e.g.][]{1995ApJS...98..617T,2000A&A...359..191G}, they do not account for
a slight overabundance of Cr relative to Fe.

The [Mn/Fe$]_{\rm obs}$ ratios cannot be fitted by any dust depletion
sequence (see Fig.~\ref{deseq}), even if non-solar values are considered for
the intrinsic [Mn/Fe$]_{\rm dc}$ ratio. The observed values can only be
recovered if one assumes possible variations of this intrinsic elemental
ratio in the range $-0.3$ to $+0.1$ dex.
%\citep[see also][]{2001A&A...370...23H}
%The five systems with [Mn/Fe$]_{\rm obs}> 0$ should be of fairly high
%metallicity since they have [Fe/H$]_{\rm obs}\la -1$. Unfortunately,
%there is no Zn measurement for three of them.
For the 17 DLA systems with both Mn and Zn measurements, there is a trend for
an increase of [Mn/Fe$]_{\rm obs}$ with increasing
[Zn/Fe$]_{\rm obs}$, detected at the $3.2\sigma$ significance level using a
Kendall rank correlation test (limits are taken into account as
true detections). This trend may be the consequence of a relation between
depletion level and metallicity. To test this suggestion, we investigate
the variations of [Zn/Fe$]_{\rm obs}$ and [Si/Fe$]_{\rm obs}$
versus [Zn/H$]_{\rm obs}$, Si being much less depleted than Mn, Cr and Fe
in the Galactic ISM. They are presented in the upper panels of
Fig.~\ref{abobs2}. For these two elemental ratios, there is a
marginal correlation with metallicity, present only for both at the
$1.7\sigma$ significance level. To confirm this possible correlation between
depletion level and metallicity would require larger samples, in particular
an extension at low metallicity, $-2.0\la [$Zn/H$]_{\rm obs}\la -1.5$.
Whereas [Cr/Fe$]_{\rm obs}$ does not show any variation with metallicity
%(as expected since Cr and Fe have comparable
%depletion factors and are produced by similar nucleosynthesis processes),
there is a clear increase of [Mn/Fe$]_{\rm obs}$ with increasing
metallicity detected at the $2.6\sigma$ significance level. This correlation
suggests for the first time that, on average, the intrinsic [Mn/Fe] ratio gets
more sub-solar toward lower metallicity, a trend which is compatible with
the pattern observed in Galactic halo stars (see Table~\ref{arcetab}).

Although the number of Ti measurements is small, the variation
of [Ti/Fe$]_{\rm obs}$ with [Zn/Fe$]_{\rm obs}$ cannot be fitted either by a
single dust depletion sequence. As for Mn, this suggests a variation of
the intrinsic [Ti/Fe$]_{\rm dc}$ ratio. Since Ti is more depleted onto
dust grains than Fe, the positive values of [Ti/Fe$]_{\rm obs}$ observed
in most cases would imply, as for the other $\alpha$-element in our study,
Si, an over-solar, intrinsic [Ti/Fe$]_{\rm dc}$ elemental ratio (see below).

\section{Discussion}\label{nucle}

\begin{figure*}
\hbox{
\psfig{figure=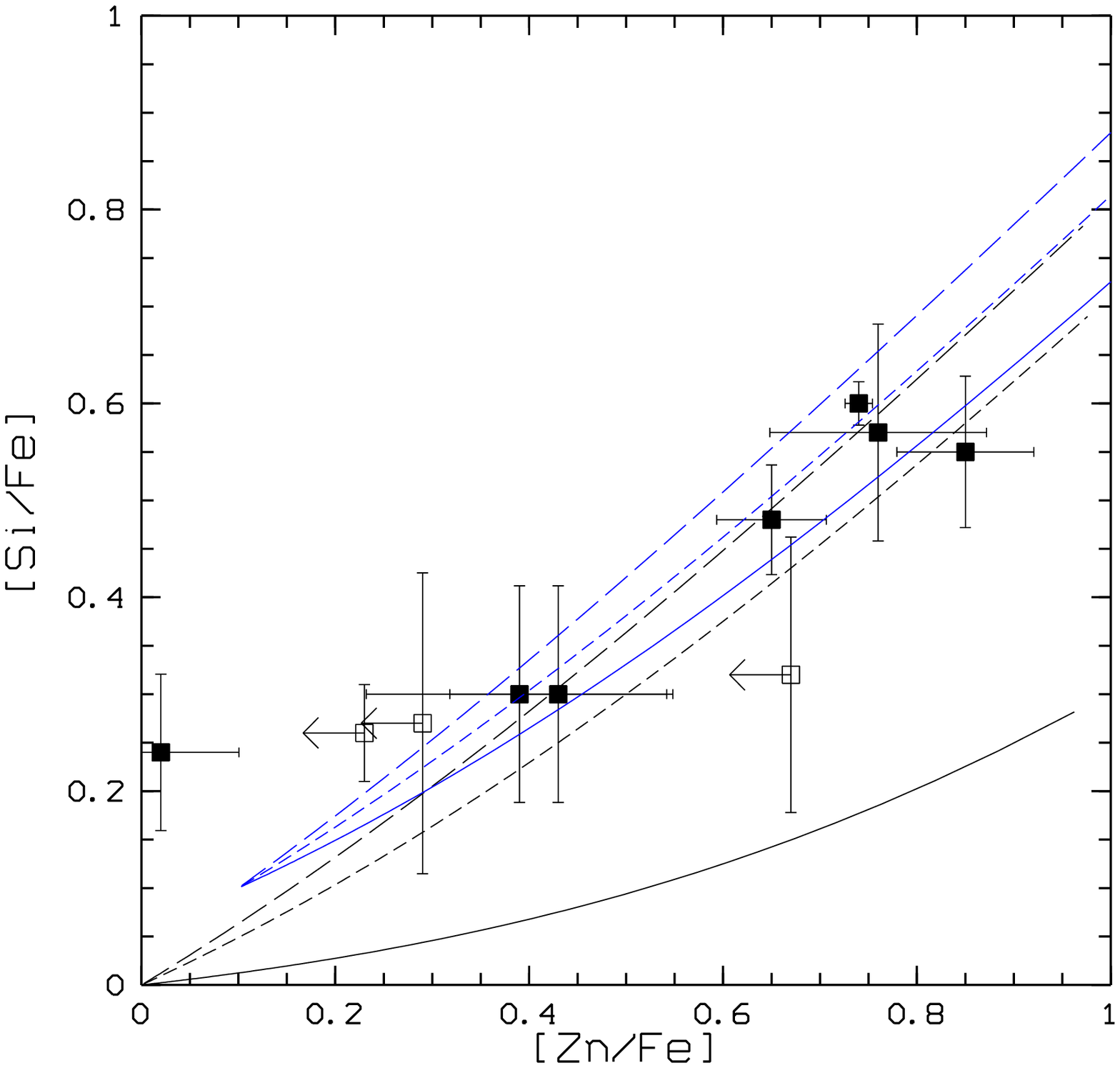,width=7.5cm,clip=,bbllx=40.pt,bblly=295.pt,bburx=555.pt,bbury=780.pt,angle=0.}\hspace{0.25cm}
\psfig{figure=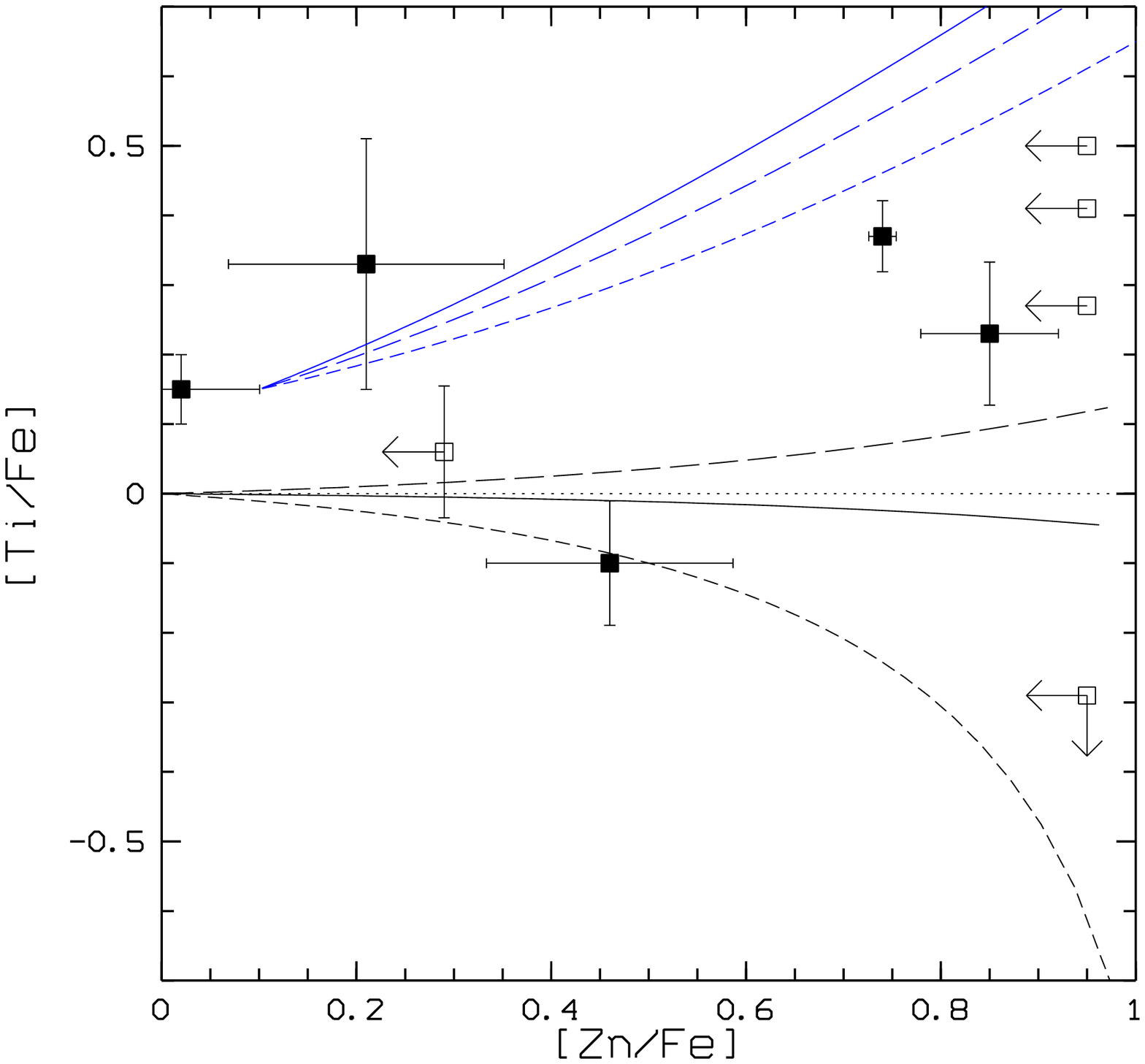,width=7.5cm,clip=,bbllx=40.pt,bblly=295.pt,bburx=555.pt,bbury=780.pt,angle=0.}}
\vspace{0.15cm}
\hbox{
\psfig{figure=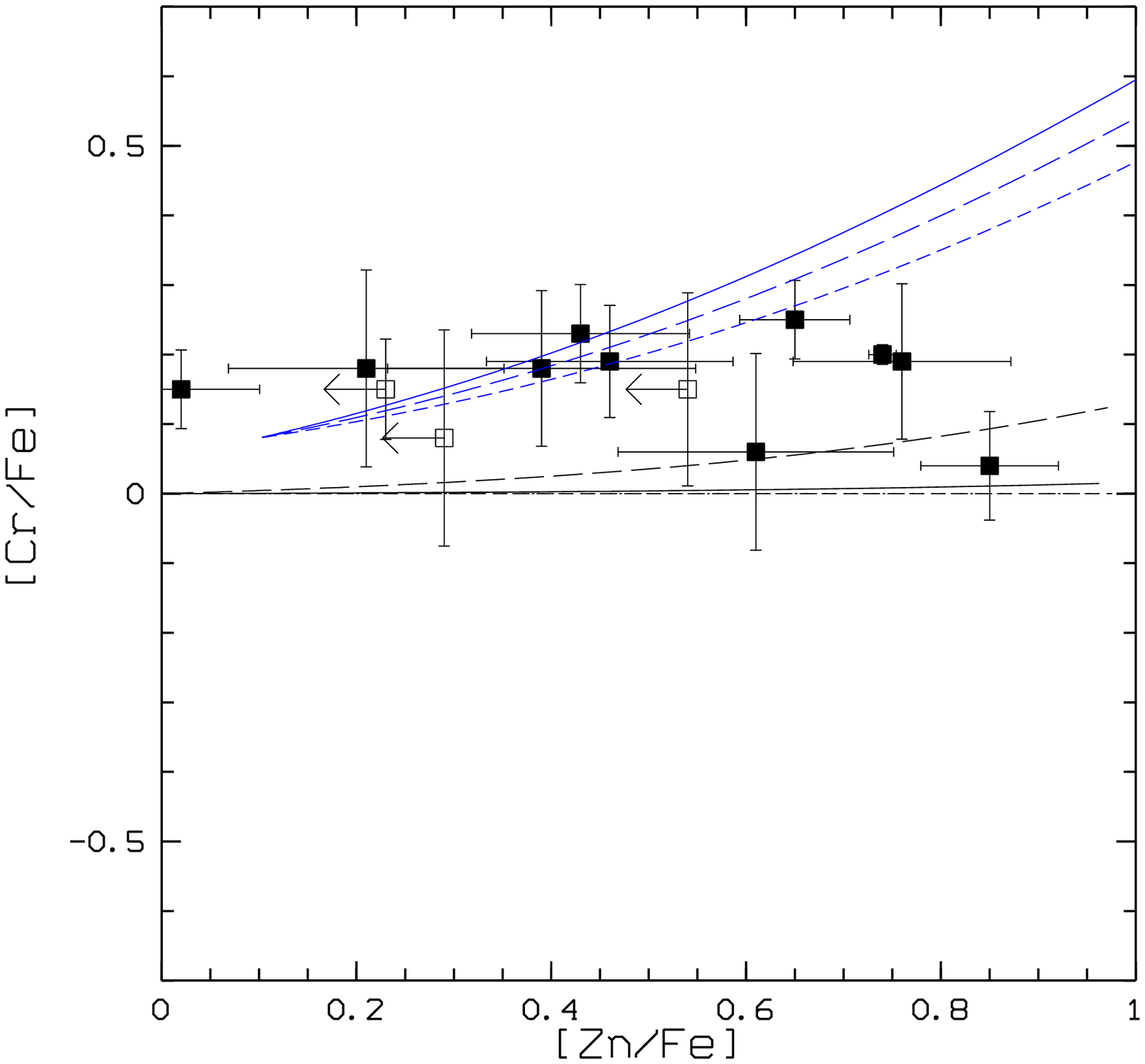,width=7.5cm,clip=,bbllx=40.pt,bblly=295.pt,bburx=555.pt,bbury=780.pt,angle=0.}\hspace{0.25cm}
\psfig{figure=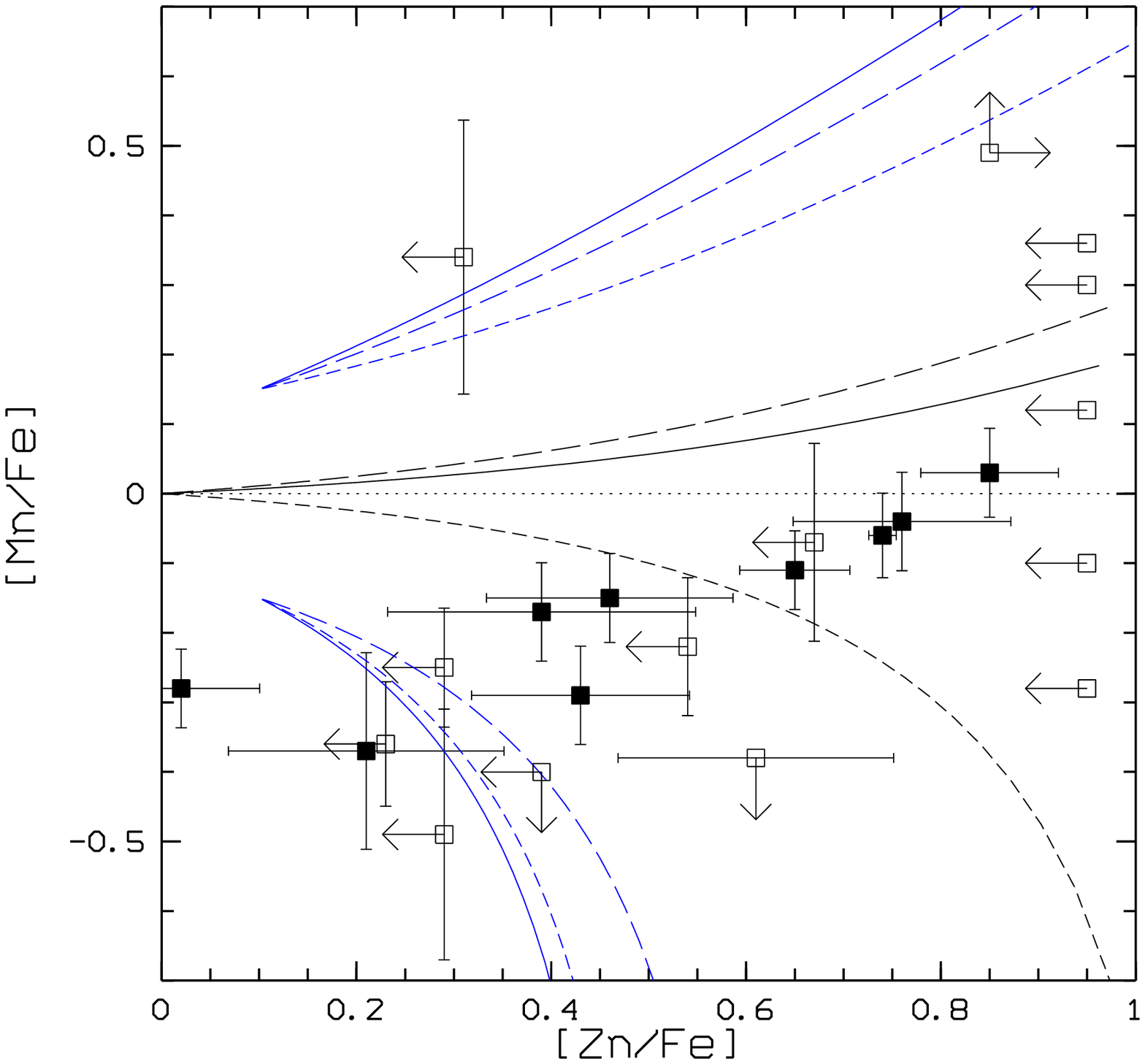,width=7.5cm,clip=,bbllx=40.pt,bblly=295.pt,bburx=555.pt,bbury=780.pt,angle=0.}}
\caption[]{Observed abundance ratios of Si, Ti, Cr and Mn relative to Fe
versus observed [Zn/Fe]. For completeness, absorbers with Ti and/or Mn but no
Zn measurements are shown assuming that [Zn/Fe$]<0.95$. Meaningful
double limits for [Ti/Fe] and [Mn/Fe] ratios are also displayed. Examples of
depletion sequence are superimposed to the data for various dust
compositions: the ISM depletion patterns of Galactic cold disk clouds
(solid lines) and Galactic warm disk (long dash) and warm halo clouds
(short dash) from \citet{1999ApJS..124..465W}. Each of these sequences
is given for intrinsically solar, over-solar and/or sub-solar ratios.
An intrinsic [Zn/Fe$]=+0.1$ ratio is typical of those observed in
Galactic halo and SMC stars (see Table~\ref{arcetab}).}
\label{deseq}
\end{figure*}

\begin{figure*}
\hbox{
\psfig{figure=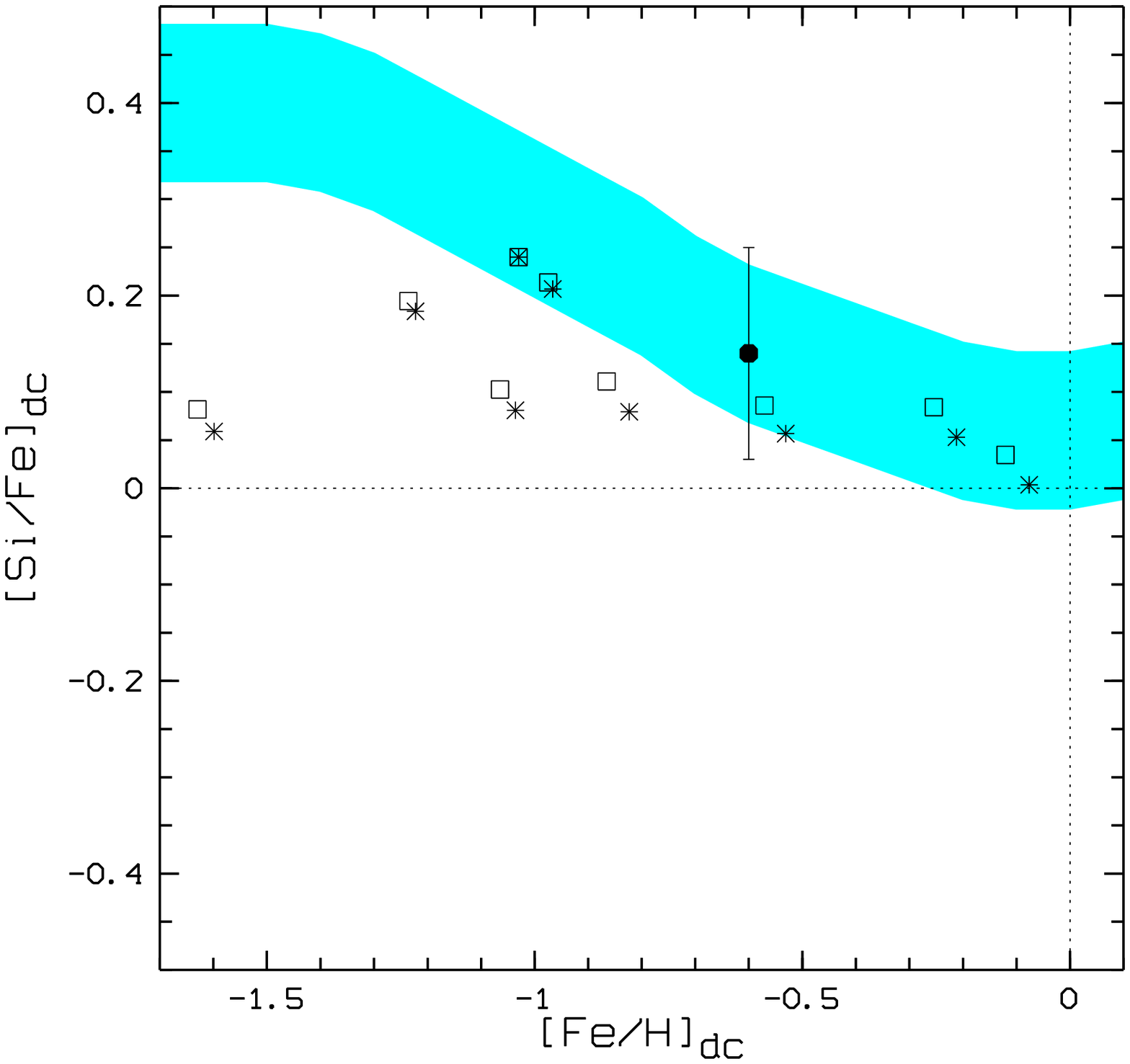,width=7.5cm,clip=,bbllx=40.pt,bblly=295.pt,bburx=555.pt,bbury=780.pt,angle=0.}\hspace{0.25cm}
\psfig{figure=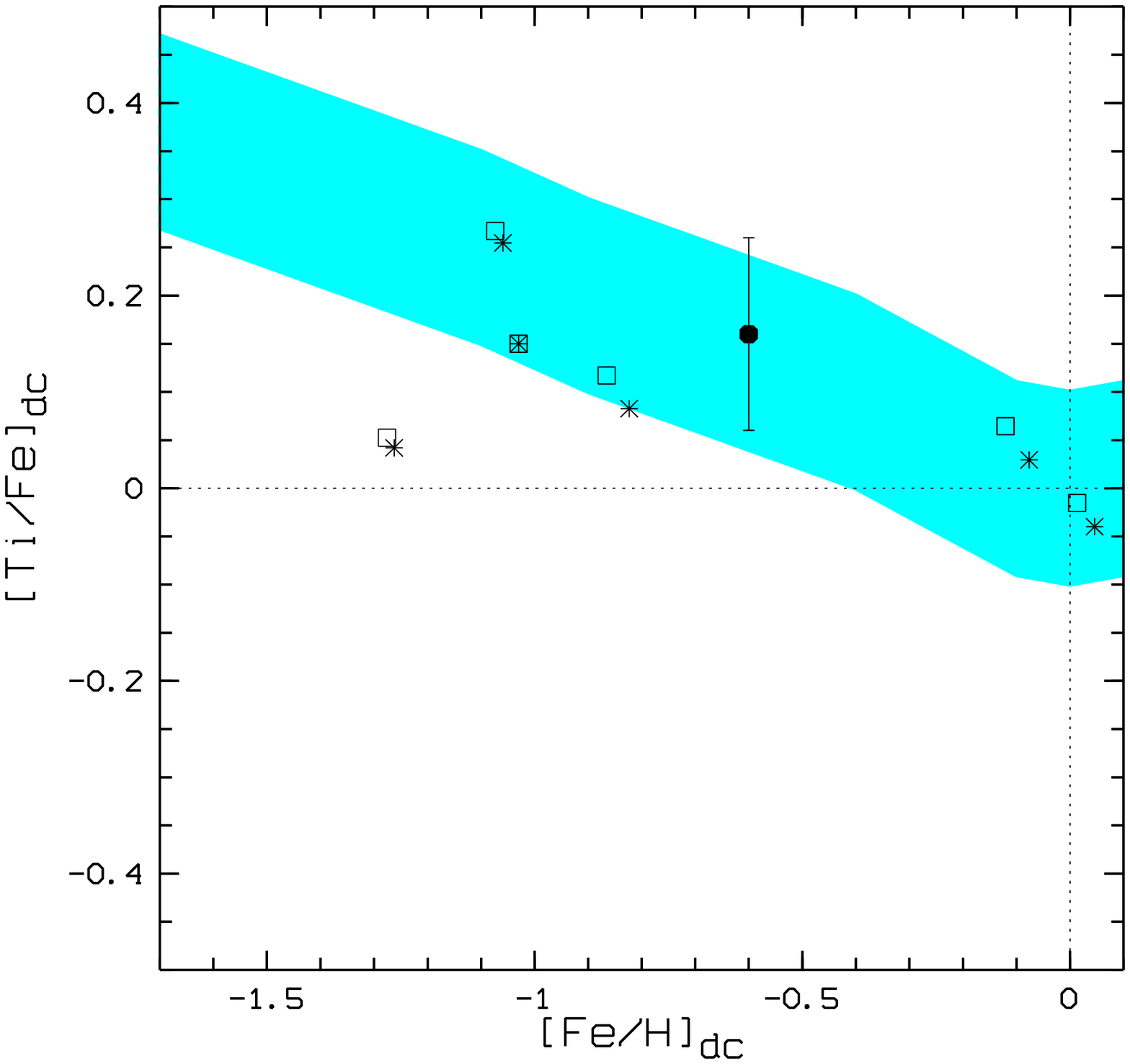,width=7.5cm,clip=,bbllx=40.pt,bblly=295.pt,bburx=555.pt,bbury=780.pt,angle=0.}}
\vspace{0.15cm}
\hbox{
\psfig{figure=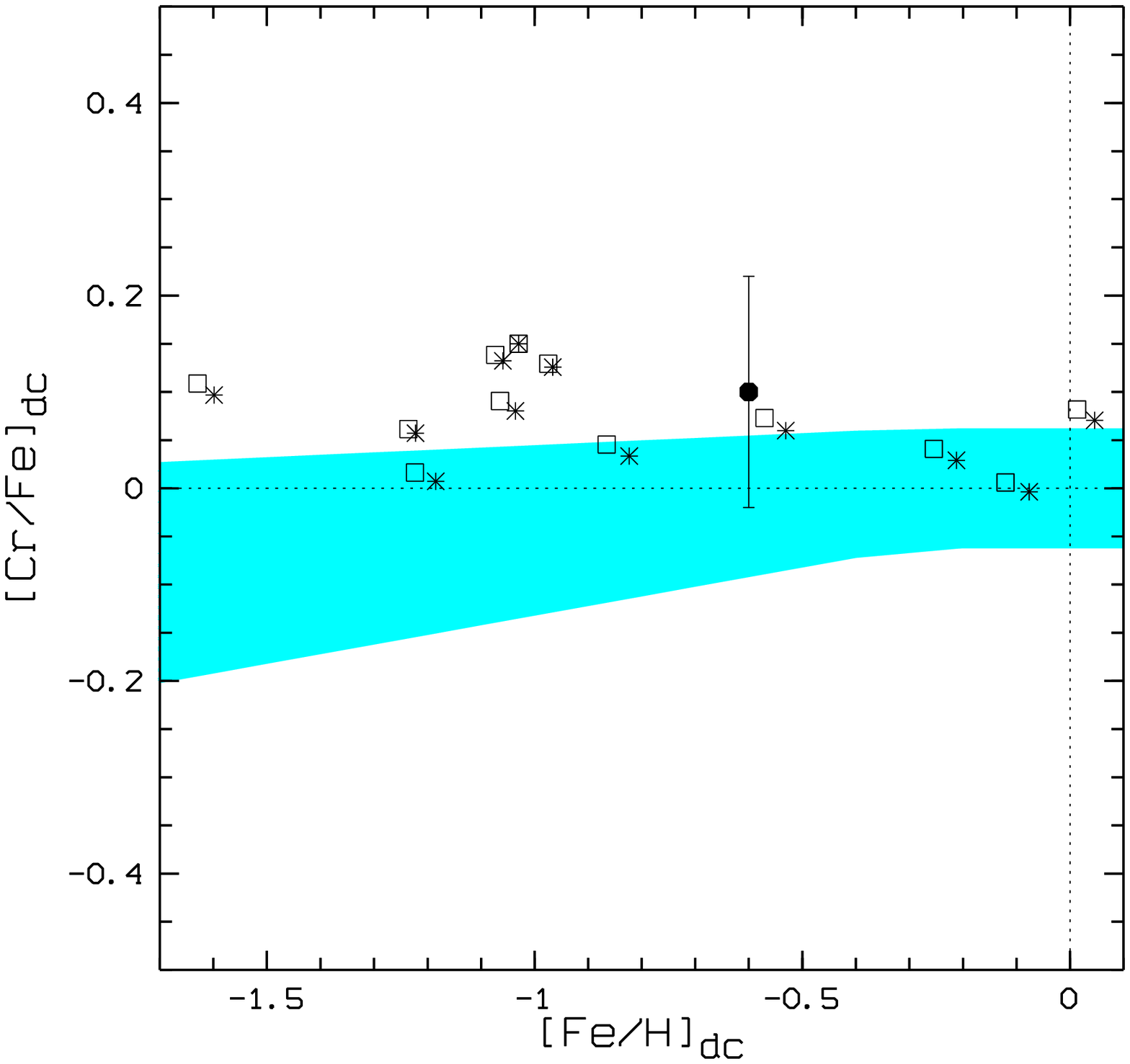,width=7.5cm,clip=,bbllx=40.pt,bblly=295.pt,bburx=555.pt,bbury=780.pt,angle=0.}\hspace{0.25cm}
\psfig{figure=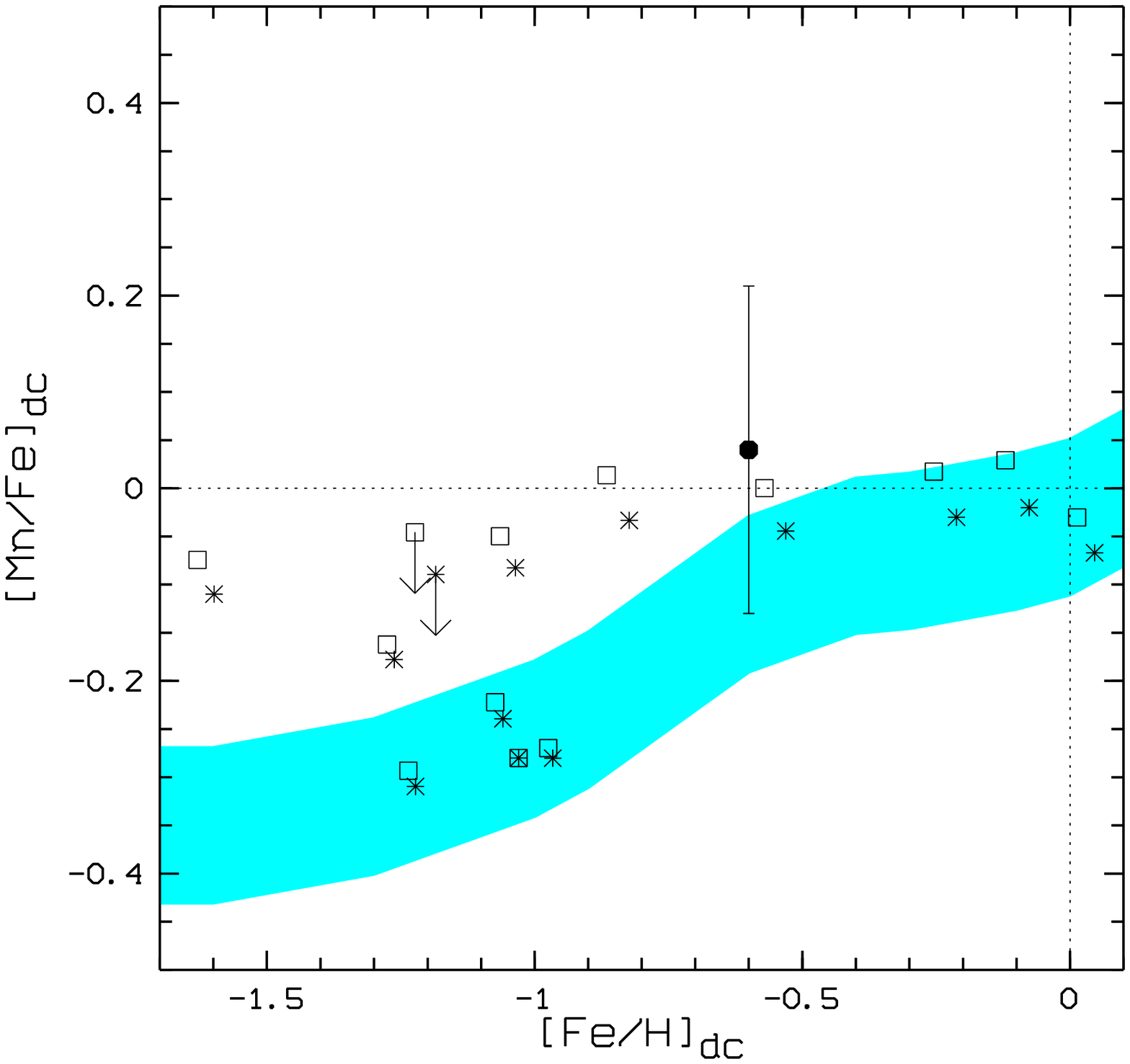,width=7.5cm,clip=,bbllx=40.pt,bblly=295.pt,bburx=555.pt,bbury=780.pt,angle=0.}}
\caption[]{Dust-corrected, i.e. intrinsic, abundance ratios of Si, Ti, Cr
and Mn relative to Fe versus dust-corrected Fe abundances, i.e. metallicities.
The observed [Zn/Fe] ratio was used as a depletion indicator with the dust
depletion patterns of Galactic warm disk (asterisks) and warm halo
clouds (squares). Errors are typically $\pm 0.1$ dex including the
uncertainty in the [Zn/Fe] ratios. The trends observed in Galactic stars
(shaded areas) are typical values for various Galactic stellar populations
and the filled circles refer to SMC stars (see Table~\ref{arcetab}).}
\label{ducor}
\end{figure*}

We now discuss the intrinsic elemental ratios derived
from Eq.~(\ref{vladilo2}) for the 24 DLA absorbers of our sample at
$0.3<z_{\rm abs}<2.2$. Since the depletion pattern of Galactic cold disk
clouds does not fit the correlation between [Si/Fe$]_{\rm obs}$
and [Zn/Fe$]_{\rm obs}$ for an intrinsic elemental
ratio [Si/Fe$]_{\rm dc}<+0.1$, we use the dust depletion patterns of
either Galactic warm disk or warm halo clouds. We present
in Fig.~\ref{ducor} the dust-corrected abundance ratios versus metallicity,
as given by [Fe/H$]_{\rm dc}$. As most of the DLA absorbers of our sample have
sub-solar abundances, we assume a value for [Zn/Fe$]_{\rm dc}$ equal to that
observed in Galactic metal-poor and SMC stars (see Table~\ref{arcetab}), i.e.
[Zn/Fe$]_{\rm dc}=+0.1$. This implies
[Si/Zn$]_{\rm dc}=[$Cr/Zn$]_{\rm dc}=0$ (see Fig.~\ref{deseq}
and Subsect.~\ref{results}). If we had assumed a solar [Zn/Fe$]_{\rm dc}$
ratio, the trends discussed below would be similar and there would be
only very small shifts ($\sim -0.05$ dex) in the absolute values of the
dust-corrected abundance ratios. 

\begin{figure*}
\hbox{
\psfig{figure=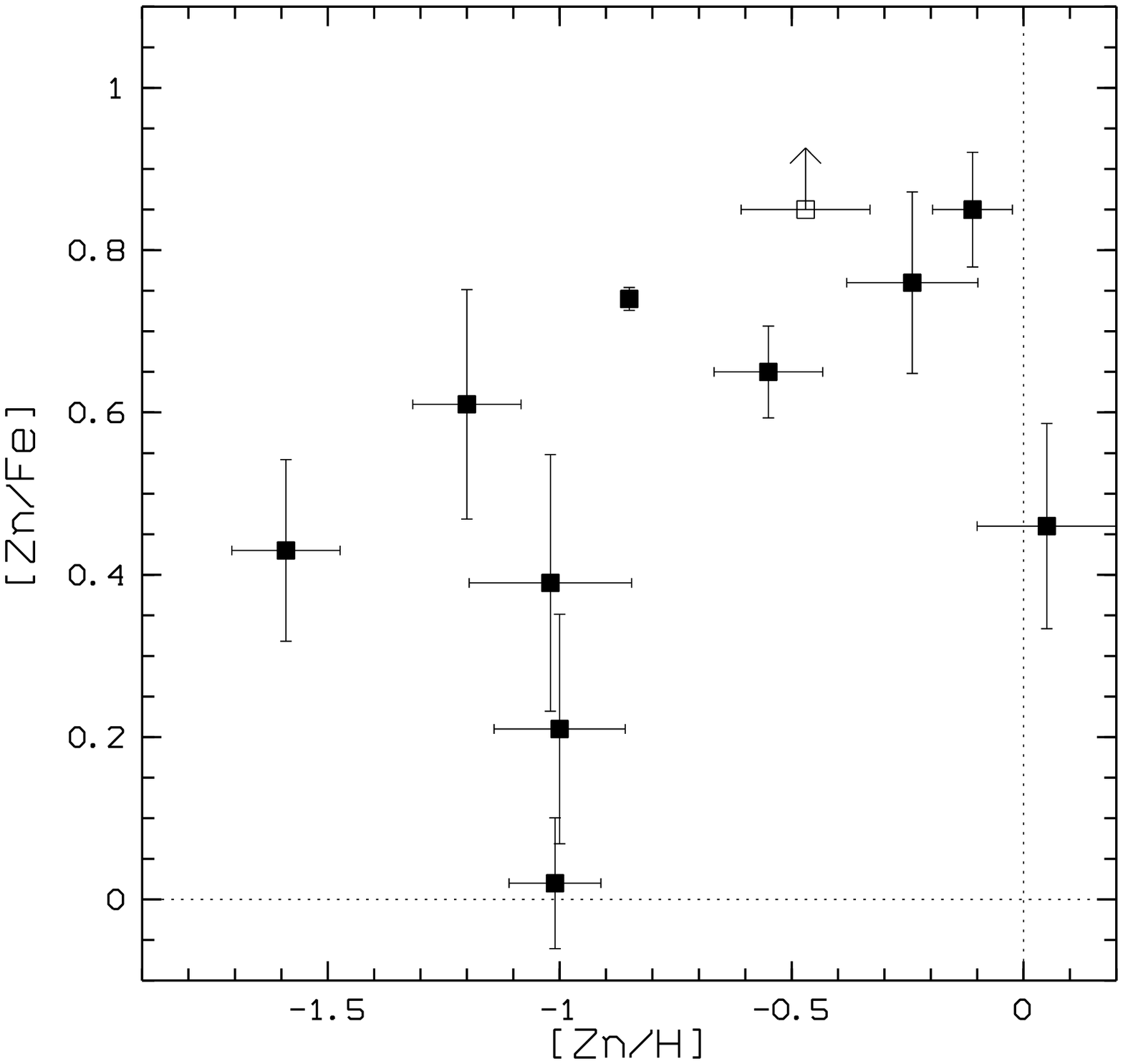,width=7.5cm,clip=,bbllx=40.pt,bblly=295.pt,bburx=555.pt,bbury=780.pt,angle=0.}\hspace{0.25cm}
\psfig{figure=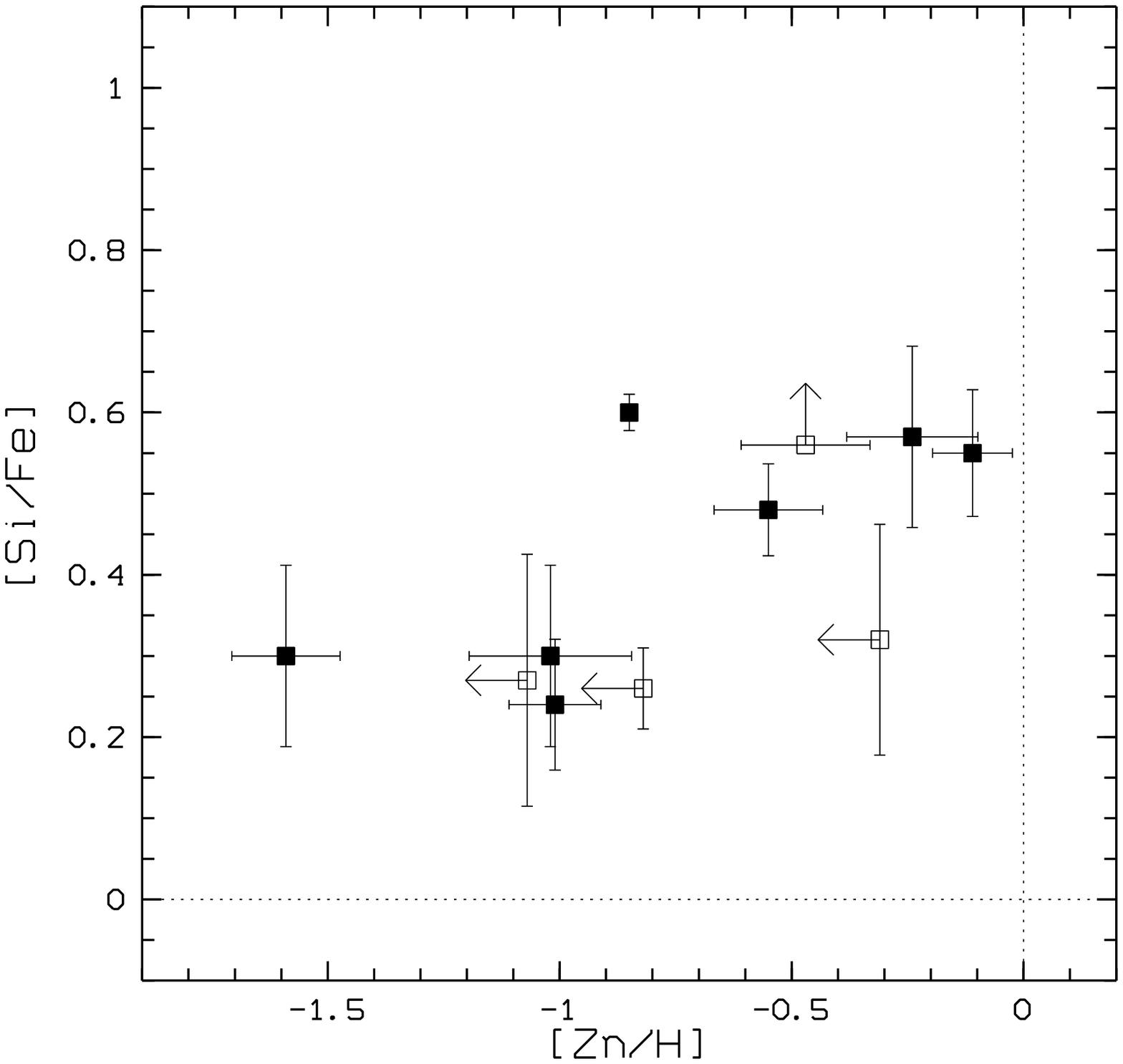,width=7.5cm,clip=,bbllx=40.pt,bblly=295.pt,bburx=555.pt,bbury=780.pt,angle=0.}}
\vspace{0.15cm}
\hbox{
\psfig{figure=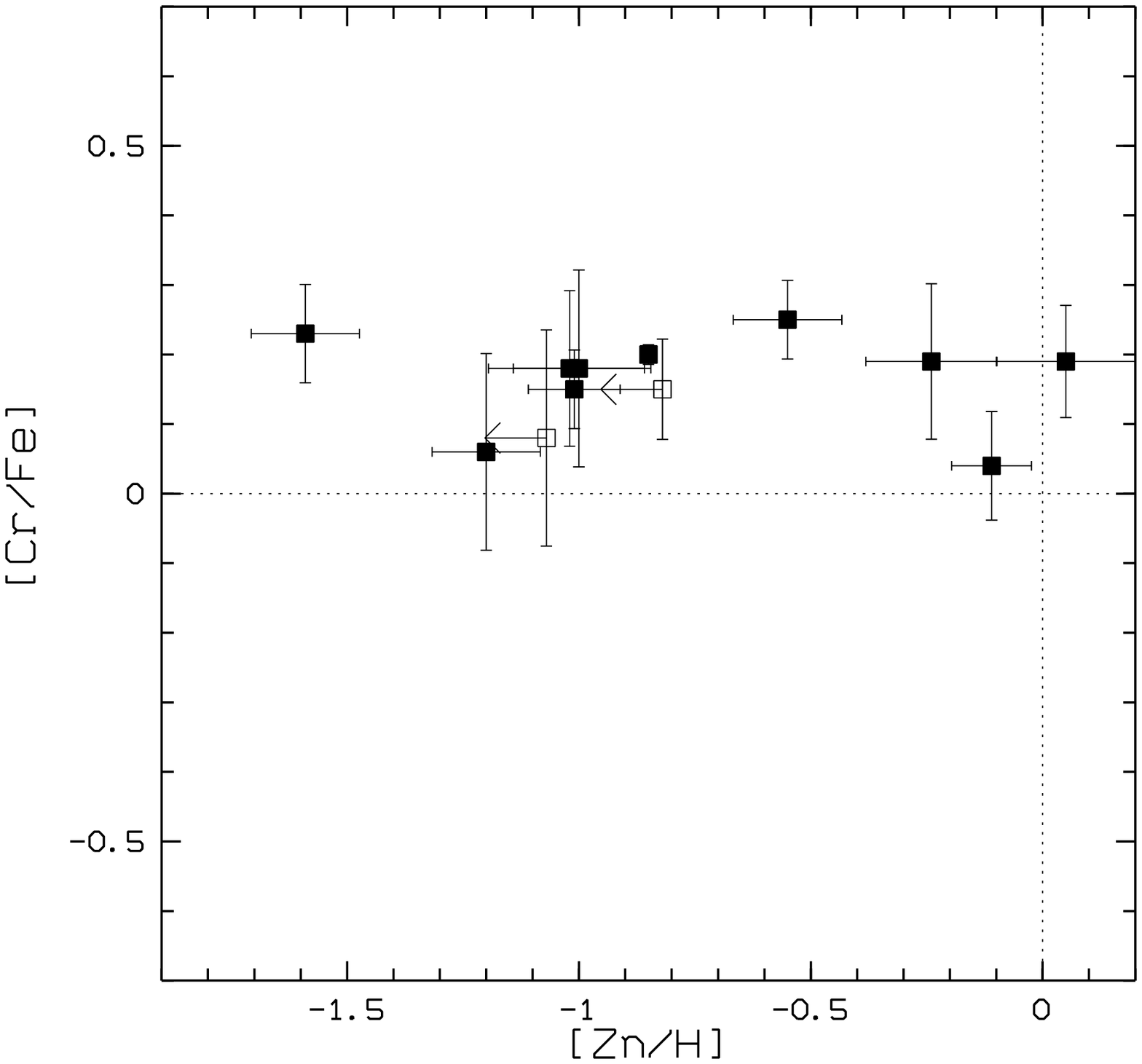,width=7.5cm,clip=,bbllx=40.pt,bblly=295.pt,bburx=555.pt,bbury=780.pt,angle=0.}\hspace{0.25cm}
\psfig{figure=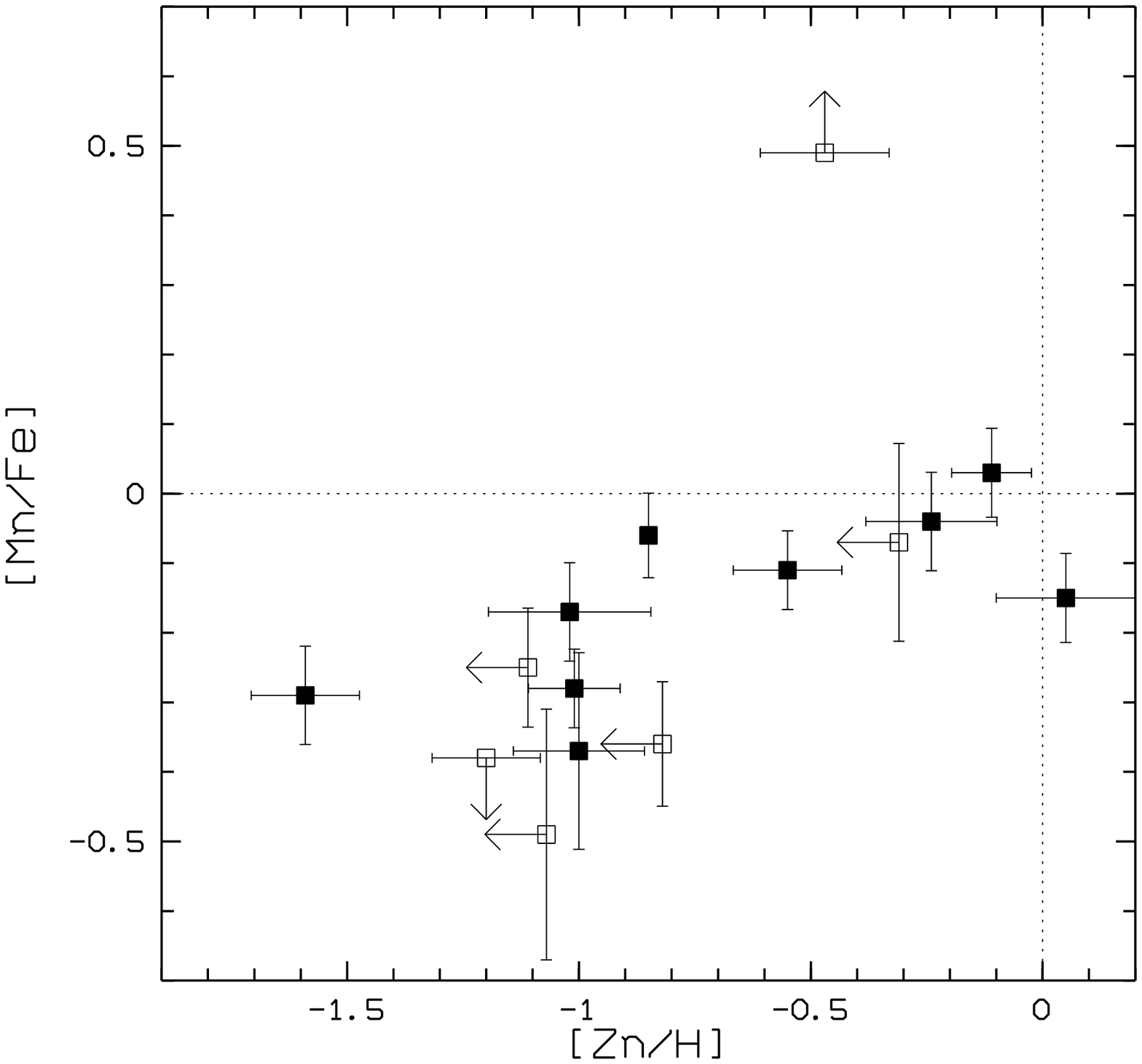,width=7.5cm,clip=,bbllx=40.pt,bblly=295.pt,bburx=555.pt,bbury=780.pt,angle=0.}}
\caption[]{Observed abundances of Zn and Si (upper panels),
Cr and Mn (lower panels) relative to Fe versus the observed abundances of Zn,
i.e. metallicities. Trends of increasing depletion level, traced by the
observed [Zn/Fe] and [Si/Fe] ratios, with metallicity are present at
the $1.7\sigma$ significance level for both ratios. The correlation
between the observed [Mn/Fe] ratio and metallicity is detected at the
$2.6\sigma$ significance level.}
\label{abobs2}
\end{figure*}

As discussed in Subsect.~\ref{results}, the importance of
nucleosynthesis effects is already revealed by the different behaviours of
the observed abundances of Mn and Cr relative to Fe. The fractions of these
two elements locked onto dust grains are similar (and also close to that for
Fe) and they have comparable solar abundances relative to Fe. However, whereas
the [Cr/Fe$]_{\rm obs}$ ratio is fairly constant whatever the metallicity, the
correlation observed between [Mn/Fe$]_{\rm obs}$ and metallicity
(Fig.~\ref{abobs2}) implies an increase of the intrinsic abundance of
Mn relative to Fe with metallicity.

To compare the intrinsic abundance patterns observed in DLA systems with those
of Galactic and SMC stars, we also give in Fig.~\ref{ducor} the
relation between the abundance ratios and metallicity observed in Galactic
thin disk and halo stars \citep[from the compilation of values
by][]{2000AJ....120.2513P} as well as the values for SMC stars \citep[from
the appendix of][]{1997ApJ...489..672W}. For the trend shown for Mn
in Galactic stars, we took into account the hyper-fine structure splitting
effects of the Mn\,{\sc i} lines \citep{2000ApJ...537L..57P} which leads to a
slightly higher [Mn/Fe] mean ratio, now equal to $-0.05$ at [Fe/H$]\ga -0.5$.

We note that the intrinsic abundances of the two $\alpha$-elements, Si and Ti,
relative to Fe show a similar trend versus metallicity, although the dust
depletion pattern used to determine these intrinsic elemental ratios was
derived from the correlation present between [Si/Fe$]_{\rm obs}$ and
[Zn/Fe$]_{\rm obs}$ only. This strengthens our choice of warm dust depletion
patterns and dust depletion correction method.

For the few DLA absorbers of fairly high
metallicity, [Fe/H$]_{\rm dc}\ga -0.5$, all four elemental
ratios: [Si/Fe$]_{\rm dc}$, [Cr/Fe$]_{\rm dc}$, [Ti/Fe$]_{\rm dc}$ and
[Mn/Fe$]_{\rm dc}$ follow the trends present in Galactic stars and, at
[Fe/H$]_{\rm dc}\simeq 0$, their abundances relative to Fe are about equal
to the solar values. The chemical evolution of these DLA systems appears
to be similar to that of our Galaxy.

At the metallicity of the SMC, [Fe/H$]\approx -0.6$, the
DLA dust-corrected abundances of Si, Ti and Cr relative to Fe, and in a lesser
extent that of Mn, are close to those of SMC stars (see Fig.~\ref{ducor}
and Table~\ref{arcetab}) with
[Si/Fe$]_{\rm dc}\sim [$Ti/Fe$]_{\rm dc}\sim [$Cr/Fe$]_{\rm dc}\sim +0.1$ and
[Mn/Fe$]_{\rm dc}\sim 0$.

However, at lower metallicity, [Fe/H$]_{\rm dc}<-0.6$, for a
substantial fraction of DLA absorbers the intrinsic abundances of Si, Ti and
Mn relative to Fe are closer to the solar values than for Galactic stars
and show little variation with decreasing metallicity. The evolution of
[Mn/Fe$]_{\rm dc}$ with metallicity is well accounted for by
the odd-even effects for the Fe-peak elements using metallicity dependent
yields \citep[for Galactic models
see e.g.][]{1995ApJS...98..617T,2000A&A...359..191G}. Furthermore, the
systems with [Si/Fe$]_{\rm dc}\sim [$Ti/Fe$]_{\rm dc}\la +0.10$ (resp.
$>+0.10$) are indeed also those with [Mn/Fe$]_{\rm dc}>-0.15$
(resp. $<-0.15$), i.e. there is an anti-correlation between [Mn/Fe$]_{\rm dc}$
and [Si/Fe$]_{\rm dc}$ or [Ti/Fe$]_{\rm dc}$. This strongly suggests that
the contribution of type Ia supernovae to the chemical enrichment of the
metal-poor DLA systems with [Si/Fe$]_{\rm dc}\sim [$Ti/Fe$]_{\rm dc}\sim +0.1$
and [Mn/Fe$]_{\rm dc}\sim -0.1$ is larger than for Galactic stars of
comparable metallicity. This conclusion is consistent with the
interpretations of \citet{1997A&A...321...45M}
and \citet{2000A&A...355..891L} based on other abundance ratios, which favor
at least for some DLA absorbers an episodic or slow star-formation scenario.
If they have experienced instantaneous starbursts, these DLA absorbers should
thus be older than 50 Myr \citep[see][]{2001ApJ...558..351M}. Since there is
a wide spread of galactic morphological types among the identified
DLA absorbers \citep{1997A&A...321..733L}, the metal-poor DLA absorbers with
abundance ratios fairly close to the solar values could indeed trace the
H\,{\sc i}-rich dwarf galaxy population.

Among the goals for future work, we identify the extension of DLA samples at
low [Zn/H] to test the possible correlation between depletion level and
metallicity, and observations of S and Zn in the same DLA systems to confirm
the lack of depletion for Zn and distinguish further between
different star-formation histories from a comparison of [S/Fe or Zn] in DLA
absorbers and in Galactic/SMC stars.

%------------------------------------------------------------------------------

\begin{acknowledgements}
We are grateful to C. W. Churchill for making available to us unpublished data
on the Ti abundances of four DLA systems and N. Prantzos for
fruitful discussion. We thank R. Srianand for the reduction of the echelle
spectrum of Q\,0453$-$423. CL acknowledges support from an ESO
postdoctoral fellowship. This work was supported by the European Community
Research and Training Network: ''The Physics of the Intergalactic Medium''.
\end{acknowledgements}

%------------------------------------------------------------------------------

\bibliographystyle{apj}
\bibliography{H3071}

\begin{thebibliography}{61}
\expandafter\ifx\csname natexlab\endcsname\relax\def\natexlab#1{#1}\fi

\bibitem[{{Ballester} {et~al.}(2000){Ballester}, {Modigliani}, {Boitquin},
  {et~al.}}]{2000Msngr.110...31B}
{Ballester}, P., {Modigliani}, A., {Boitquin}, O., {et~al.} 2000, The
  Messenger, 101, 31

\bibitem[{{Bergeron} \& {D'Odorico}(1986)}]{1986MNRAS.220..833B}
{Bergeron}, J. \& {D'Odorico}, S. 1986, MNRAS, 220, 833

\bibitem[{{Bergeron} \& {Stasi\'nska}(1986)}]{1986A&A...169....1B}
{Bergeron}, J. \& {Stasi\'nska}, G. 1986, A\&A, 169, 1

\bibitem[{{Bergeson} {et~al.}(1996){Bergeson}, {Mullman}, {Wickliffe},
  {et~al.}}]{1996ApJ...464.1044B}
{Bergeson}, S.~D., {Mullman}, K.~L., {Wickliffe}, M.~E., {et~al.} 1996, ApJ,
  464, 1044

\bibitem[{{Blades} {et~al.}(1982){Blades}, {Hunstead}, {Murdoch}, \&
  {Pettini}}]{1982MNRAS.200.1091B}
{Blades}, J.~C., {Hunstead}, R.~W., {Murdoch}, H.~S., \& {Pettini}, M. 1982,
  MNRAS, 200, 1091

\bibitem[{{Blades} {et~al.}(1985){Blades}, {Hunstead}, {Murdoch}, \&
  {Pettini}}]{1985ApJ...288..580B}
---. 1985, ApJ, 288, 580

\bibitem[{{Boiss\'e} {et~al.}(1998){Boiss\'e}, {Le Brun}, {Bergeron}, \&
  {Deharveng}}]{1998A&A...333..841B}
{Boiss\'e}, P., {Le Brun}, V., {Bergeron}, J., \& {Deharveng}, J.-M. 1998,
  A\&A, 333, 841

\bibitem[{{Churchill} {et~al.}(2000{\natexlab{a}}){Churchill}, {Mellon},
  {Charlton}, {et~al.}}]{2000ApJS..130...91C}
{Churchill}, C.~W., {Mellon}, R.~R., {Charlton}, J.~C., {et~al.}
  2000{\natexlab{a}}, ApJS, 130, 91

\bibitem[{{Churchill} {et~al.}(2000{\natexlab{b}}){Churchill}, {Mellon},
  {Charlton}, {et~al.}}]{2000ApJ...543..577C}
---. 2000{\natexlab{b}}, ApJ, 543, 577

\bibitem[{{Cohen} {et~al.}(1994){Cohen}, {Barlow}, {Beaver},
  {et~al.}}]{1994ApJ...421..453C}
{Cohen}, R.~D., {Barlow}, T.~A., {Beaver}, E.~A., {et~al.} 1994, ApJ, 421, 453

\bibitem[{{de la Varga} {et~al.}(2000){de la Varga}, {Reimers}, {Tytler},
  {Barlow}, \& {Burles}}]{2000A&A...363...69D}
{de la Varga}, A., {Reimers}, D., {Tytler}, D., {Barlow}, T.~A., \& {Burles},
  S. 2000, A\&A, 363, 69

\bibitem[{{Fontana} \& {Ballester}(1995)}]{1995Msngr..80...37F}
{Fontana}, A. \& {Ballester}, P. 1995, The Messenger, 80, 37

\bibitem[{{Goswami} \& {Prantzos}(2000)}]{2000A&A...359..191G}
{Goswami}, A. \& {Prantzos}, N. 2000, A\&A, 359, 191

\bibitem[{{Hoffman} {et~al.}(1996){Hoffman}, {Woosley}, {Fuller}, \&
  {Meyer}}]{1996ApJ...460..478H}
{Hoffman}, R.~D., {Woosley}, S.~E., {Fuller}, G.~M., \& {Meyer}, B.~S. 1996,
  ApJ, 460, 478

\bibitem[{{Kulkarni} {et~al.}(1997){Kulkarni}, {Fall}, \&
  {Truran}}]{1997ApJ...484L...7K}
{Kulkarni}, V.~P., {Fall}, S.~M., \& {Truran}, J.~W. 1997, ApJ, 484, L7

\bibitem[{{Lanzetta} \& {Bowen}(1992)}]{1992ApJ...391...48L}
{Lanzetta}, K.~M. \& {Bowen}, D.~V. 1992, ApJ, 391, 48

\bibitem[{{Le Brun} {et~al.}(1997){Le Brun}, {Bergeron}, {Boiss\'e}, \&
  {Deharveng}}]{1997A&A...321..733L}
{Le Brun}, V., {Bergeron}, J., {Boiss\'e}, P., \& {Deharveng}, J.-M. 1997,
  A\&A, 321, 733

\bibitem[{{Ledoux} {et~al.}(1998){Ledoux}, {Petitjean}, {Bergeron}, {Wampler},
  \& {Srianand}}]{1998A&A...337...51L}
{Ledoux}, C., {Petitjean}, P., {Bergeron}, J., {Wampler}, E.~J., \& {Srianand},
  R. 1998, A\&A, 337, 51

\bibitem[{{Ledoux} {et~al.}(2002){Ledoux}, {Srianand}, \&
  {Petitjean}}]{2002A&Asubmitted..L}
{Ledoux}, C., {Srianand}, R., \& {Petitjean}, P. 2002, MNRAS, submitted

\bibitem[{{Legrand} {et~al.}(2000){Legrand}, {Kunth}, {Roy}, {Mas-Hesse}, \&
  {Walsh}}]{2000A&A...355..891L}
{Legrand}, F., {Kunth}, D., {Roy}, J.-R., {Mas-Hesse}, J.~M., \& {Walsh}, J.~R.
  2000, A\&A, 355, 891

\bibitem[{{Lehner} {et~al.}(2001){Lehner}, {Sembach}, {Dufton}, {Rolleston}, \&
  {Keenan}}]{2001ApJ...551..781L}
{Lehner}, N., {Sembach}, K.~R., {Dufton}, P.~L., {Rolleston}, W. R.~J., \&
  {Keenan}, F.~P. 2001, ApJ, 551, 781

\bibitem[{{Lopez} {et~al.}(1999){Lopez}, {Reimers}, {Rauch}, {Sargent}, \&
  {Smette}}]{1999ApJ...513..598L}
{Lopez}, S., {Reimers}, D., {Rauch}, M., {Sargent}, W. L.~W., \& {Smette}, A.
  1999, ApJ, 513, 598

\bibitem[{{Lu} {et~al.}(1996){Lu}, {Sargent}, {Barlow}, {Churchill}, \&
  {Vogt}}]{1996ApJS..107..475L}
{Lu}, L., {Sargent}, W. L.~W., {Barlow}, T.~A., {Churchill}, C.~W., \& {Vogt},
  S.~S. 1996, ApJS, 107, 475

\bibitem[{{Lu} {et~al.}(1995){Lu}, {Savage}, {Tripp}, \&
  {Meyer}}]{1995ApJ...447..597L}
{Lu}, L., {Savage}, B.~D., {Tripp}, T.~M., \& {Meyer}, D.~M. 1995, ApJ, 447,
  597

\bibitem[{{Matteucci} {et~al.}(1997){Matteucci}, {Molaro}, \&
  {Vladilo}}]{1997A&A...321...45M}
{Matteucci}, F., {Molaro}, P., \& {Vladilo}, G. 1997, A\&A, 321, 45

\bibitem[{{Matteucci} \& {Recchi}(2001)}]{2001ApJ...558..351M}
{Matteucci}, F. \& {Recchi}, S. 2001, ApJ, 558, 351

\bibitem[{{McWilliam}(1997)}]{1997ARA&A..35..503W}
{McWilliam}, A. 1997, ARA\&A, 35, 503

\bibitem[{{McWilliam} {et~al.}(1995){McWilliam}, {Preston}, {Sneden}, \&
  {Searle}}]{1995AJ....109.2757M}
{McWilliam}, A., {Preston}, G.~W., {Sneden}, C., \& {Searle}, L. 1995, AJ, 109,
  2757

\bibitem[{{Meyer} {et~al.}(1995){Meyer}, {Lanzetta}, \&
  {Wolfe}}]{1995ApJ...451L..13M}
{Meyer}, D.~M., {Lanzetta}, K.~M., \& {Wolfe}, A.~M. 1995, ApJ, 451, L13

\bibitem[{{Meyer} \& {York}(1992)}]{1992ApJ...399L.121M}
{Meyer}, D.~M. \& {York}, D.~G. 1992, ApJ, 399, L121

\bibitem[{{Morton}(1991)}]{1991ApJS...77..119M}
{Morton}, D.~C. 1991, ApJS, 77, 119

\bibitem[{{Nagataki}(1999)}]{1999ApJ...511..341N}
{Nagataki}, S. 1999, ApJ, 511, 341

\bibitem[{{Petitjean} \& {Bergeron}(1994)}]{1994A&A...283..759P}
{Petitjean}, P. \& {Bergeron}, J. 1994, A\&A, 283, 759

\bibitem[{{Petitjean} {et~al.}(2002){Petitjean}, {Srianand}, \&
  {Ledoux}}]{2002MNRASsubmitted..P}
{Petitjean}, P., {Srianand}, R., \& {Ledoux}, C. 2002, MNRAS, in press,
  astro-ph/0201477

\bibitem[{{Pettini} {et~al.}(1999){Pettini}, {Ellison}, {Steidel}, \&
  {Bowen}}]{1999ApJ...510..576P}
{Pettini}, M., {Ellison}, S.~L., {Steidel}, C.~C., \& {Bowen}, D.~V. 1999, ApJ,
  510, 576

\bibitem[{{Pettini} {et~al.}(2000){Pettini}, {Ellison}, {Steidel}, {Shapley},
  \& {Bowen}}]{2000ApJ...532...65P}
{Pettini}, M., {Ellison}, S.~L., {Steidel}, C.~C., {Shapley}, A.~E., \&
  {Bowen}, D.~V. 2000, ApJ, 532, 65

\bibitem[{{Pettini} {et~al.}(1997{\natexlab{a}}){Pettini}, {King}, {Smith}, \&
  {Hunstead}}]{1997ApJ...478..536P}
{Pettini}, M., {King}, D.~L., {Smith}, L.~J., \& {Hunstead}, R.~W.
  1997{\natexlab{a}}, ApJ, 478, 536

\bibitem[{{Pettini} {et~al.}(1994){Pettini}, {Smith}, {Hunstead}, \&
  {King}}]{1994ApJ...426...79P}
{Pettini}, M., {Smith}, L.~J., {Hunstead}, R.~W., \& {King}, D.~L. 1994, ApJ,
  426, 79

\bibitem[{{Pettini} {et~al.}(1997{\natexlab{b}}){Pettini}, {Smith}, {King}, \&
  {Hunstead}}]{1997ApJ...486..665P}
{Pettini}, M., {Smith}, L.~J., {King}, D.~L., \& {Hunstead}, R.~W.
  1997{\natexlab{b}}, ApJ, 486, 665

\bibitem[{{Prochaska} \& {McWilliam}(2000)}]{2000ApJ...537L..57P}
{Prochaska}, J.~X. \& {McWilliam}, A. 2000, ApJ, 537, L57

\bibitem[{{Prochaska} {et~al.}(2000){Prochaska}, {Naumov}, {Carney},
  {McWilliam}, \& {Wolfe}}]{2000AJ....120.2513P}
{Prochaska}, J.~X., {Naumov}, S.~O., {Carney}, B.~W., {McWilliam}, A., \&
  {Wolfe}, A.~M. 2000, AJ, 120, 2513

\bibitem[{{Prochaska} \& {Wolfe}(1997{\natexlab{a}})}]{1997ApJ...487...73P}
{Prochaska}, J.~X. \& {Wolfe}, A.~M. 1997{\natexlab{a}}, ApJ, 487, 73

\bibitem[{{Prochaska} \& {Wolfe}(1997{\natexlab{b}})}]{1997ApJ...474..140P}
---. 1997{\natexlab{b}}, ApJ, 474, 140

\bibitem[{{Prochaska} \& {Wolfe}(1999)}]{1999ApJS..121..369P}
---. 1999, ApJS, 121, 369

\bibitem[{{Prochaska} {et~al.}(2001){Prochaska}, {Wolfe}, {Tytler},
  {et~al.}}]{2001ApJS..137...21P}
{Prochaska}, J.~X., {Wolfe}, A.~M., {Tytler}, D., {et~al.} 2001, ApJS, 137, 21

\bibitem[{{Rao} \& {Turnshek}(2000)}]{2000ApJS..130....1R}
{Rao}, S.~M. \& {Turnshek}, D.~A. 2000, ApJS, 130, 1

\bibitem[{{Ryan} {et~al.}(1996){Ryan}, {Norris}, \&
  {Beers}}]{1996ApJ...471.254R}
{Ryan}, S.~G., {Norris}, J.~E., \& {Beers}, T.~C. 1996, ApJ, 471, 254

\bibitem[{{Savage} \& {Sembach}(1996)}]{1996ARA&A..34..279S}
{Savage}, B.~D. \& {Sembach}, K.~R. 1996, ARA\&A, 34, 279

\bibitem[{{Steidel} {et~al.}(1997){Steidel}, {Dickinson}, {Meyer},
  {Adelberger}, \& {Sembach}}]{1997ApJ...480..568S}
{Steidel}, C.~C., {Dickinson}, M., {Meyer}, D.~M., {Adelberger}, K.~L., \&
  {Sembach}, K.~R. 1997, ApJ, 480, 568

\bibitem[{{Thielemann} {et~al.}(1996){Thielemann}, {Nomoto}, \&
  {Hashimoto}}]{1996ApJ...460..408T}
{Thielemann}, F.-K., {Nomoto}, K., \& {Hashimoto}, M. 1996, ApJ, 460, 408

\bibitem[{{Timmes} {et~al.}(1995){Timmes}, {Woosley}, \&
  {Weaver}}]{1995ApJS...98..617T}
{Timmes}, F.~X., {Woosley}, S.~E., \& {Weaver}, T.~A. 1995, ApJS, 98, 617

\bibitem[{{Umeda} \& {Nomoto}(2002)}]{2002ApJ...565..385U}
{Umeda}, H. \& {Nomoto}, K. 2002, ApJ, 565, 385

\bibitem[{{V\'eron-Cetty} \& {V\'eron}(2001)}]{2001A&A...374...92V}
{V\'eron-Cetty}, M.-P. \& {V\'eron}, P. 2001, A\&A, 374, 92

\bibitem[{{Viegas}(1995)}]{1995MNRAS.276..268V}
{Viegas}, S.~M. 1995, MNRAS, 276, 268

\bibitem[{{Vladilo}(1998)}]{1998ApJ...493..583V}
{Vladilo}, G. 1998, ApJ, 493, 583

\bibitem[{{Vladilo} {et~al.}(2001){Vladilo}, {Centuri\'on}, {Bonifacio}, \&
  {Howk}}]{2001ApJ...557.1007V}
{Vladilo}, G., {Centuri\'on}, M., {Bonifacio}, P., \& {Howk}, J.~C. 2001, ApJ,
  557, 1007

\bibitem[{{Vladilo} {et~al.}(1997){Vladilo}, {Centuri\'on}, {Falomo}, \&
  {Molaro}}]{1997A&A...327...47V}
{Vladilo}, G., {Centuri\'on}, M., {Falomo}, R., \& {Molaro}, P. 1997, A\&A,
  327, 47

\bibitem[{{Welty} {et~al.}(1999){Welty}, {Hobbs}, {Lauroesch},
  {et~al.}}]{1999ApJS..124..465W}
{Welty}, D.~E., {Hobbs}, L.~M., {Lauroesch}, J.~T., {et~al.} 1999, ApJS, 124,
  465

\bibitem[{{Welty} {et~al.}(1997){Welty}, {Lauroesch}, {Blades}, {Hobbs}, \&
  {York}}]{1997ApJ...489..672W}
{Welty}, D.~E., {Lauroesch}, J.~T., {Blades}, J.~C., {Hobbs}, L.~M., \& {York},
  D.~G. 1997, ApJ, 489, 672

\bibitem[{{Welty} {et~al.}(2001){Welty}, {Lauroesch}, {Blades}, {Hobbs}, \&
  {York}}]{2001ApJ...554L..75W}
---. 2001, ApJ, 554, L75

\bibitem[{{Wolfe} {et~al.}(1986){Wolfe}, {Turnshek}, {Smith}, \&
  {Cohen}}]{1986ApJS...61..249W}
{Wolfe}, A.~M., {Turnshek}, D.~A., {Smith}, H.~E., \& {Cohen}, R.~D. 1986,
  ApJS, 61, 249

\end{thebibliography}

%\listofobjects

%------------------------------------------------------------------------------

\begin{table*}
\caption[]{Observational data:}
\begin{flushleft}
\begin{tabular}{lllllcccl}
\hline
Quasar & mag$^{\rm a}$ & $z^{\rm a}_{\rm em}$ & Telescope & Instrument & Wavelength   & Resolution  & Exposure & Comment\\
       &               &                      &           &            & range (\AA ) & FWHM (\AA ) & time (s) &     \\
% UT Date 1997/09/23--26 1995/11/18--21 1997/09/23--26
% HST release date: 1994/11/02 1997/07/13
\hline
0058$+$019 & 17.16\,(V) & 1.959 & ESO 3.6 m & CASPEC & 3510--4797\phantom{$^{\rm b}$} & 0.16 &           21600 &          \\
           &            &       & HST       & FOS    & 1572--2311\phantom{$^{\rm b}$} & 0.84 & \phantom{0}1590 & G190H    \\
0453$-$423 & 17.06\,(V) & 2.661 & ESO 3.6 m & CASPEC & 3680--5096\phantom{$^{\rm b}$} & 0.16 &           16200 &          \\
           &            &       & HST       & FOS    & 1572--2311\phantom{$^{\rm b}$} & 0.84 & \phantom{0}1700 & G190H    \\
           &            &       & HST       & FOS    & 2223--3277\phantom{$^{\rm b}$} & 1.19 & \phantom{0}2440 & G270H    \\
1122$-$168 & 16.20\,(R) & 2.400 & VLT-UT2   & UVES   & 3350--3875\phantom{$^{\rm b}$} & 0.10 &           26400 & Dic\,\#1 \\
           &            &       & VLT-UT2   & UVES   & 4780--6815$^{\rm b}$           & 0.10 &           26400 & Dic\,\#1 \\
           &            &       & VLT-UT2   & UVES   & 3875--4780\phantom{$^{\rm b}$} & 0.10 &           27000 &          \\
2128$-$123 & 16.11\,(V) & 0.501 & ESO 3.6 m & CASPEC & 3510--4797\phantom{$^{\rm b}$} & 0.16 &           21600 &          \\
           &            &       & HST       & FOS    & 1572--2311\phantom{$^{\rm b}$} & 0.84 & \phantom{0}5221 & G190H    \\
\hline
\end{tabular}
\end{flushleft}
$^{\rm a}$ Apparent magnitudes and emission redshifts are
from \citet{2001A&A...374...92V}.\\
$^{\rm b}$ The small 5765--5837 \AA\ interval was not covered due to the
gap between red CCDs.
\label{obsetab}
\end{table*}

%------------------------------------------------------------------------------

\begin{table*}
\caption[]{Model fit parameters for individual components:}
\hbox{\begin{tabular}{llllll}
\hline
n & $z_{\rm abs}$ & $\Delta V$    & Ion & $b$ ($\sigma _b$) & $\log N$($\sigma _{\log N}$)\\
  &               & km s$^{-1}$   &     & km s$^{-1}$       &                             \\
\hline
\multicolumn{2}{c}{Q\,0058$+$019}&&&&\\
\hline
%SiII, TiII, ZnII, CrII out of the range
1 &\phantom{$-$}0.612102 &\phantom{0}$-$51 & Mg\,{\sc i} &\phantom{0}7.0 {\scriptsize (....)}&          $<$11.00$^{\rm a}$          \\
  &                      &                 & Mg\,{\sc ii}&                                   &\phantom{$>$}12.73{\scriptsize (0.05)}\\
  &                      &                 & Mn\,{\sc ii}&                                   &          $<$11.98$^{\rm a}$          \\
  &                      &                 & Fe\,{\sc ii}&                                   &\phantom{$>$}12.79{\scriptsize (0.07)}\\
2 &\phantom{$-$}0.612377 &\phantom{$-$00}0 & Mg\,{\sc i} &          13.0 {\scriptsize (....)}&\phantom{$>$}12.35{\scriptsize (0.05)}\\
  &                      &                 & Mg\,{\sc ii}&                                   &          $>$15.98                    \\
  &                      &                 & Mn\,{\sc ii}&                                   &\phantom{$>$}12.88{\scriptsize (0.04)}\\
  &                      &                 & Fe\,{\sc ii}&                                   &\phantom{$>$}15.11{\scriptsize (0.08)}\\
3 &\phantom{$-$}0.612519 &\phantom{0}$+$26 & Mg\,{\sc i} &          13.0 {\scriptsize (....)}&\phantom{$>$}12.13{\scriptsize (0.06)}\\
  &                      &                 & Mg\,{\sc ii}&                                   &          $>$14.18                    \\
  &                      &                 & Mn\,{\sc ii}&                                   &\phantom{$>$}12.65{\scriptsize (0.06)}\\
  &                      &                 & Fe\,{\sc ii}&                                   &\phantom{$>$}14.98{\scriptsize (0.09)}\\
4 &\phantom{$-$}0.612719 &\phantom{0}$+$64 & Mg\,{\sc i} &          13.0 {\scriptsize (....)}&\phantom{$>$}11.81{\scriptsize (0.07)}\\
  &                      &                 & Mg\,{\sc ii}&                                   &          $>$15.01                    \\
  &                      &                 & Mn\,{\sc ii}&                                   &\phantom{$>$}12.10{\scriptsize (0.16)}\\
  &                      &                 & Fe\,{\sc ii}&                                   &\phantom{$>$}13.99{\scriptsize (0.06)}\\
5 &\phantom{$-$}0.612973 &          $+$111 & Mg\,{\sc i} &          11.2 {\scriptsize (0.7)} &\phantom{$>$}11.46{\scriptsize (0.13)}\\
  &                      &                 & Mg\,{\sc ii}&                                   &          $>$13.89                    \\
  &                      &                 & Mn\,{\sc ii}&                                   &          $<$11.98$^{\rm a}$          \\
  &                      &                 & Fe\,{\sc ii}&                                   &\phantom{$>$}13.75{\scriptsize (0.04)}\\
\hline
\multicolumn{2}{c}{Q\,0453$-$423}&&&&\\
\hline
%SiII, ZnII: out of the range, CrII very bad in the blue, TiII: out of the range
1 &\phantom{$-$}0.725853 &\phantom{0}$-$33 & Mg\,{\sc i} &          26.5 {\scriptsize (0.8)} &\phantom{$>$}12.16{\scriptsize (0.04)}\\
  &                      &                 & Mg\,{\sc ii}&                                   &          $>$13.97                    \\
  &                      &                 & Mn\,{\sc ii}&                                   &\phantom{$>$}12.34{\scriptsize (0.06)}\\
  &                      &                 & Fe\,{\sc ii}&                                   &\phantom{$>$}14.27{\scriptsize (0.03)}\\
2 &\phantom{$-$}0.726043 &\phantom{$-$00}0 & Mg\,{\sc i} &          16.2 {\scriptsize (1.0)} &\phantom{$>$}12.21{\scriptsize (0.03)}\\
  &                      &                 & Mg\,{\sc ii}&                                   &          $>$13.53                    \\
  &                      &                 & Mn\,{\sc ii}&                                   &\phantom{$>$}12.22{\scriptsize (0.07)}\\
  &                      &                 & Fe\,{\sc ii}&                                   &\phantom{$>$}14.01{\scriptsize (0.05)}\\
3 &\phantom{$-$}0.726280 &\phantom{0}$+$41 & Mg\,{\sc i} &          23.4 {\scriptsize (0.9)} &\phantom{$>$}11.56{\scriptsize (0.09)}\\
  &                      &                 & Mg\,{\sc ii}&                                   &          $>$13.57                    \\
  &                      &                 & Mn\,{\sc ii}&                                   &\phantom{$>$}11.95{\scriptsize (0.11)}\\
  &                      &                 & Fe\,{\sc ii}&                                   &\phantom{$>$}13.75{\scriptsize (0.02)}\\
\hline
%MnII from P&B94
1 &\phantom{$-$}1.148673 &          $-$129 & Fe\,{\sc ii}&          30.0 {\scriptsize (....)}&\phantom{$>$}13.95{\scriptsize (0.09)}\\
2 &\phantom{$-$}1.149060 &\phantom{0}$-$75 & Si\,{\sc ii}&          30.0 {\scriptsize (....)}&        $\le$14.72{\scriptsize (0.13)}\\
  &                      &                 & Ti\,{\sc ii}&                                   &          $<$12.62$^{\rm a}$          \\
  &                      &                 & Cr\,{\sc ii}&                                   &        $\le$12.98{\scriptsize (0.08)}\\
  &                      &                 & Fe\,{\sc ii}&                                   &\phantom{$>$}14.55{\scriptsize (0.08)}\\
  &                      &                 & Zn\,{\sc ii}&                                   &        $\le$12.01{\scriptsize (0.21)}\\
3 &\phantom{$-$}1.149598 &\phantom{$-$00}0 & Si\,{\sc ii}&          23.5 {\scriptsize (3.3)} &        $\le$14.97{\scriptsize (0.07)}\\
  &                      &                 & Ti\,{\sc ii}&                                   &          $<$12.62$^{\rm a}$          \\
  &                      &                 & Cr\,{\sc ii}&                                   &        $\le$12.85{\scriptsize (0.10)}\\
  &                      &                 & Fe\,{\sc ii}&                                   &\phantom{$>$}14.83{\scriptsize (0.06)}\\
  &                      &                 & Zn\,{\sc ii}&                                   &        $\le$12.27{\scriptsize (0.11)}\\
4 &\phantom{$-$}1.150024 &\phantom{0}$+$59 & Fe\,{\sc ii}&          25.9 {\scriptsize (3.1)} &\phantom{$>$}14.28{\scriptsize (0.08)}\\
5 &\phantom{$-$}1.150482 &          $+$123 & Fe\,{\sc ii}&          16.3 {\scriptsize (1.4)} &\phantom{$>$}13.55{\scriptsize (0.03)}\\
6 &\phantom{$-$}1.150980 &          $+$193 & Si\,{\sc ii}&          30.0 {\scriptsize (....)}&        $\le$14.71{\scriptsize (0.12)}\\
  &                      &                 & Fe\,{\sc ii}&                                   &\phantom{$>$}13.71{\scriptsize (0.03)}\\
7 &\phantom{$-$}1.151371 &          $+$247 & Si\,{\sc ii}&          30.0 {\scriptsize (....)}&          $<$14.47$^{\rm a}$          \\
  &                      &                 & Fe\,{\sc ii}&                                   &\phantom{$>$}13.67{\scriptsize (0.03)}\\
8 &\phantom{$-$}1.151953 &          $+$329 & Si\,{\sc ii}&          20.3 {\scriptsize (0.9)} &        $\le$15.69{\scriptsize (0.02)}\\
  &                      &                 & Fe\,{\sc ii}&                                   &\phantom{$>$}13.88{\scriptsize (0.02)}\\

\hline
\multicolumn{6}{l}{$^{\rm a}$ 3$\sigma$ upper limits.}\\
\end{tabular}
\begin{tabular}{llllll}
\hline
n & $z_{\rm abs}$ & $\Delta V$    & Ion & $b$ ($\sigma _b$) & $\log N$($\sigma _{\log N}$)\\
  &               & km s$^{-1}$   &     & km s$^{-1}$       &                             \\
\hline
\multicolumn{2}{c}{Q\,1122$-$168}&&&&\\
\hline
%SiII out of the range, CrII triplet completely blended
1 &\phantom{$-$}0.681309 &\phantom{0}$-$99 & Mg\,{\sc i} &\phantom{0}4.0 {\scriptsize (0.2)} &\phantom{$>$}10.75{\scriptsize (0.07)}\\
  &                      &                 & Mg\,{\sc ii}&                                   &\phantom{$>$}12.77{\scriptsize (0.02)}\\
  &                      &                 & Fe\,{\sc ii}&                                   &\phantom{$>$}12.39{\scriptsize (0.02)}\\
2 &\phantom{$-$}0.681415 &\phantom{0}$-$80 & Mg\,{\sc i} &\phantom{0}8.6 {\scriptsize (1.0)} &\phantom{$>$}10.38{\scriptsize (0.20)}\\
  &                      &                 & Mg\,{\sc ii}&                                   &\phantom{$>$}11.88{\scriptsize (0.04)}\\
  &                      &                 & Fe\,{\sc ii}&                                   &\phantom{$>$}11.45{\scriptsize (0.13)}\\
3 &\phantom{$-$}0.681862 &\phantom{$-$00}0 & Mg\,{\sc i} &\phantom{0}3.4 {\scriptsize (0.1)} &\phantom{$>$}10.82{\scriptsize (0.06)}\\
  &                      &                 & Ca\,{\sc ii}&                                   &\phantom{$>$}10.69{\scriptsize (0.14)}\\
  &                      &                 & Ti\,{\sc ii}&                                   &\phantom{$>$}11.22{\scriptsize (0.08)}\\
  &                      &                 & Mn\,{\sc ii}&                                   &\phantom{$>$}11.52{\scriptsize (0.07)}\\
  &                      &                 & Fe\,{\sc ii}&                                   &\phantom{$>$}13.86{\scriptsize (0.03)}\\
  &                      &                 & Zn\,{\sc ii}&                                   &          $<$11.26$^{\rm a}$          \\
4 &\phantom{$-$}0.681919 &\phantom{0}$+$10 & Mg\,{\sc i} &\phantom{0}8.2 {\scriptsize (0.3)} &\phantom{$>$}11.01{\scriptsize (0.05)}\\
  &                      &                 & Ca\,{\sc ii}&                                   &\phantom{$>$}10.59{\scriptsize (0.23)}\\
  &                      &                 & Ti\,{\sc ii}&                                   &\phantom{$>$}11.39{\scriptsize (0.07)}\\
  &                      &                 & Mn\,{\sc ii}&                                   &\phantom{$>$}11.66{\scriptsize (0.07)}\\
  &                      &                 & Fe\,{\sc ii}&                                   &\phantom{$>$}13.79{\scriptsize (0.02)}\\
  &                      &                 & Zn\,{\sc ii}&                                   &          $<$11.26$^{\rm a}$          \\
5 &\phantom{$-$}0.682025 &\phantom{0}$+$29 & Mg\,{\sc i} &\phantom{0}8.9 {\scriptsize (0.4)} &\phantom{$>$}10.91{\scriptsize (0.07)}\\
  &                      &                 & Ca\,{\sc ii}&                                   &\phantom{$>$}10.46{\scriptsize (0.30)}\\
  &                      &                 & Fe\,{\sc ii}&                                   &\phantom{$>$}13.67{\scriptsize (0.02)}\\
6 &\phantom{$-$}0.682202 &\phantom{0}$+$61 & Mg\,{\sc i} &          17.8 {\scriptsize (0.6)} &\phantom{$>$}11.54{\scriptsize (0.02)}\\
  &                      &                 & Ca\,{\sc ii}&                                   &\phantom{$>$}11.16{\scriptsize (0.08)}\\
  &                      &                 & Ti\,{\sc ii}&                                   &        $\le$11.30{\scriptsize (0.12)}\\
  &                      &                 & Mn\,{\sc ii}&                                   &        $\le$11.76{\scriptsize (0.07)}\\
  &                      &                 & Fe\,{\sc ii}&                                   &\phantom{$>$}13.91{\scriptsize (0.01)}\\
7 &\phantom{$-$}0.682353 &\phantom{0}$+$88 & Mg\,{\sc i} &\phantom{0}7.3 {\scriptsize (0.4)} &\phantom{$>$}10.90{\scriptsize (0.07)}\\
  &                      &                 & Ca\,{\sc ii}&                                   &\phantom{$>$}10.61{\scriptsize (0.20)}\\
  &                      &                 & Fe\,{\sc ii}&                                   &\phantom{$>$}13.31{\scriptsize (0.03)}\\
8 &\phantom{$-$}0.682459 &          $+$106 & Mg\,{\sc i }&          11.7 {\scriptsize (1.6)} &\phantom{$>$}10.79{\scriptsize (0.10)}\\
  &                      &                 & Ca\,{\sc ii}&                                   &\phantom{$>$}10.83{\scriptsize (0.15)}\\
  &                      &                 & Fe\,{\sc ii}&                                   &\phantom{$>$}13.07{\scriptsize (0.05)}\\
9 &\phantom{$-$}0.682560 &          $+$125 & Mg\,{\sc i} &\phantom{0}7.7 {\scriptsize (0.1)} &\phantom{$>$}11.69{\scriptsize (0.01)}\\
  &                      &                 & Ca\,{\sc ii}&                                   &\phantom{$>$}11.41{\scriptsize (0.04)}\\
  &                      &                 & Ti\,{\sc ii}&                                   &        $\le$11.11{\scriptsize (0.12)}\\
  &                      &                 & Mn\,{\sc ii}&                                   &        $\le$11.35{\scriptsize (0.12)}\\
  &                      &                 & Fe\,{\sc ii}&                                   &\phantom{$>$}13.73{\scriptsize (0.01)}\\
  &                      &                 & Zn\,{\sc ii}&                                   &          $<$11.26$^{\rm a}$          \\
10&\phantom{$-$}0.682693 &          $+$148 & Mg\,{\sc i} &          10.6 {\scriptsize (0.3)} &\phantom{$>$}11.30{\scriptsize (0.03)}\\
  &                      &                 & Ca\,{\sc ii}&                                   &\phantom{$>$}10.92{\scriptsize (0.11)}\\
  &                      &                 & Fe\,{\sc ii}&                                   &\phantom{$>$}12.98{\scriptsize (0.01)}\\
11&\phantom{$-$}0.682928 &          $+$190 & Mg\,{\sc i} &\phantom{0}5.2 {\scriptsize (0.2)} &\phantom{$>$}10.77{\scriptsize (0.07)}\\
  &                      &                 & Ca\,{\sc ii}&                                   &\phantom{$>$}10.73{\scriptsize (0.13)}\\
  &                      &                 & Fe\,{\sc ii}&                                   &\phantom{$>$}12.81{\scriptsize (0.01)}\\
\hline
\multicolumn{2}{c}{Q\,2128$-$123}&&&&\\
\hline
%SiII, ZnII, CrII out of the range, limit on TiII from 3242
%1 &          $-$0.000055 &\phantom{0}$-$17 & Ca\,{\sc ii}&\phantom{$>$}12.46 & 0.02 & 10.7 & 0.7 \\
1 &\phantom{$-$}0.429637 &\phantom{0}$-$37 & Mg\,{\sc i} &          11.2 {\scriptsize (0.9)} &          $<$10.88$^{\rm a}$          \\
  &                      &                 & Mg\,{\sc ii}&                                   &\phantom{$>$}12.46{\scriptsize (0.02)}\\
%  &                      &                 & Ti\,{\sc ii}&                                   &          $<$11.69$^{\rm a}$          \\
  &                      &                 & Mn\,{\sc ii}&                                   &          $<$11.77$^{\rm a}$          \\
  &                      &                 & Fe\,{\sc ii}&                                   &\phantom{$>$}12.66{\scriptsize (0.05)}\\
2 &\phantom{$-$}0.429815 &\phantom{$-$00}0 & Mg\,{\sc i} &\phantom{0}8.3 {\scriptsize (0.4)} &\phantom{$>$}12.18{\scriptsize (0.02)}\\
  &                      &                 & Mg\,{\sc ii}&                                   &          $>$14.11                    \\
%  &                      &                 & Ti\,{\sc ii}&                                   &          $<$11.69$^{\rm a}$          \\
  &                      &                 & Mn\,{\sc ii}&                                   &        $\le$12.07{\scriptsize (0.08)}\\
  &                      &                 & Fe\,{\sc ii}&                                   &          $>$14.08                    \\
\hline
\end{tabular}}
\label{paratab}
\end{table*}

%-----------------------------------------------------------------------------

\begin{table*}
\caption[]{Observed abundance ratios for DLA systems at $0.3<z_{\rm abs}<2.2$:}
%\scriptsize{
\begin{tabular}{lccc@{~}lcccccl}
\hline
Quasar & $z_{\rm abs}$ & $\log N($H\,{\sc i}$)$ & [Fe/H]$^{\rm ~2}$ & & [Zn/Fe] & [Cr/Fe] & [Mn/Fe] & [Si/Fe] & [Ti/Fe] & Refs. \\%FWHM  %nb of main comp.
       &               &                        &                  & &         &         &         &         &         &       \\%(\AA )%
\hline
0058$+$019 & 0.612 & 20.14{\scriptsize (0.09)}   &  $-$0.41{\scriptsize (0.10)}& &  $+$0.46{\scriptsize (0.13)}  &  $+$0.19{\scriptsize (0.08)} &      $-$0.15{\scriptsize (0.06)} &  ...                         &      $-$0.10{\scriptsize (0.09)} & $\star$,a,b\\%0.16,0.10
0118$-$272 & 0.558 & ...$^{\rm ~1}$              &  $-$1.06                    & &  ...                          &  ...                         &      $+$0.36{\scriptsize (0.07)} &  ...                         &      $+$0.50{\scriptsize (0.07)} & c          \\%0.20
0215$+$015 & 1.345 & 19.57{\scriptsize (0.20)}   &  $-$0.80{\scriptsize (0.23)}& &  ...                          &  ...                         &      $+$0.30{\scriptsize (0.17)} & $>$$+$0.32                   &  ...                             & d,e        \\%0.37
0216$+$080 & 1.769 & 20.00{\scriptsize (0.18)}   &  $-$0.98{\scriptsize (0.20)}& &  $<$$+$0.67                   &  $<$$+$0.36                  &      $-$0.07{\scriptsize (0.14)} &  $+$0.32{\scriptsize (0.14)} &  ...                             & f          \\
0302$-$223 & 1.009 & 20.36{\scriptsize (0.11)}   &  $-$1.20{\scriptsize (0.12)}& &  $+$0.65{\scriptsize (0.06)}  &  $+$0.25{\scriptsize (0.06)} &      $-$0.11{\scriptsize (0.06)} &  $+$0.48{\scriptsize (0.06)} &  ...                             & a          \\%0.10
0449$-$134 & 1.267 & ...$^{\rm ~1}$              &  $-$0.38                    & &  ...                          &  ...                         &      $-$0.28{\scriptsize (0.04)} &  ...                         &  ...                             & f          \\%0.10
0450$-$132 & 1.174 & ...$^{\rm ~1}$              &  $-$0.39                    & & $<$$+$0.54                    &  $+$0.15{\scriptsize (0.14)} &      $-$0.22{\scriptsize (0.10)} &  ...                         &  ...                             & f          \\%0.10
0453$-$423 & 0.726 & ...$^{\rm ~1}$              &  $-$0.97                    & &  ...                          &  ...                         &      $+$0.12{\scriptsize (0.09)} &  ...                         &  ...                             & $\star$    \\%0.16
0453$-$423 & 1.150 & ...$^{\rm ~1}$              &  $-$0.33                    & & $\le$$+$0.31                  & $\le$$+$0.00                 &      $+$0.34{\scriptsize (0.20)} & $\le$$+$0.11                 &   $<$$+$0.59                     & $\star$,g  \\%0.16,0.65
0454$+$039 & 0.860 & 20.69{\scriptsize (0.07)}   &  $-$1.03{\scriptsize (0.09)}& &  $+$0.02{\scriptsize (0.08)}  &  $+$0.15{\scriptsize (0.06)} &      $-$0.28{\scriptsize (0.06)} &  $+$0.24{\scriptsize (0.08)} &      $+$0.15{\scriptsize (0.05)} & a,b        \\%0.10
0528$-$250 & 2.141 & 20.70{\scriptsize (0.08)}   &  $-$1.36{\scriptsize (0.17)}& &  $<$$+$0.29                   &  $+$0.08{\scriptsize (0.16)} &      $-$0.49{\scriptsize (0.18)} &  $+$0.27{\scriptsize (0.16)} &  ...                             & f,h        \\%
0551$-$366 & 1.962 & 20.50{\scriptsize (0.07)}   &  $-$0.96{\scriptsize (0.11)}&$^{\rm 3}$& $+$0.85{\scriptsize (0.07)}   &  $+$0.04{\scriptsize (0.08)} &      $+$0.03{\scriptsize (0.06)} & $+$0.55{\scriptsize (0.08)} & $+$0.23{\scriptsize (0.10)}      & i        \\%
%0809$+$483 & 0.437 & 20.80{\scriptsize (0.20)}  &  $>$$-$1.47                 & &  ...                          &  ...                         &   $<$$+$0.61                     &  ...                         &  ...                             & f,l        \\%1.00
0935$+$417 & 1.373 & 20.52{\scriptsize (0.10)}   &  $-$1.21{\scriptsize (0.14)}& &  $+$0.21{\scriptsize (0.14)}  &  $+$0.18{\scriptsize (0.14)} &      $-$0.37{\scriptsize (0.14)} &  ...                         &      $+$0.33{\scriptsize (0.18)} & j,k        \\%0.35
1104$-$181 & 1.662 & 20.85{\scriptsize (0.01)}   &  $-$1.59{\scriptsize (0.02)}&$^{\rm 4}$& $+$0.74{\scriptsize (0.01)}   &  $+$0.20{\scriptsize (0.01)} & $-$0.06{\scriptsize (0.06)}      & $+$0.60{\scriptsize (0.02)} & $+$0.37{\scriptsize (0.05)}      & l        \\%0.10/0.9
1122$-$168 & 0.682 & 20.45{\scriptsize (0.05)}   &  $-$1.40{\scriptsize (0.05)}&$^{\rm 5}$& $<$$+$0.29                    &  ...                         & $-$0.25{\scriptsize (0.09)}      &  ...                        & $+$0.06{\scriptsize (0.09)}      & $\star$,m\\%0.10
           &       &                             &                             &$^{\rm 6}$& $<$$+$0.39                    &  ...                         & $\le$$-$0.40{\scriptsize (0.12)} &  ...                        & $\le$$-$0.04{\scriptsize (0.12)} & $\star$  \\
%1209$+$107 & 0.629 & 20.20{\scriptsize (0.10)}  &  $-$0.89{\scriptsize (0.13)}& &  ...                          &  ...                         &   $<$$+$0.57                     &  ...                         &  ...                             & m,n        \\%2.50
%NOTE FOR THIS ONE, NUMBER OF COMPONENTS SAMPLED IN FEII ARE LIKELY TO BE MORE THAN ONE, AND THEREFORE N(FEII) CANNOT BE COMPARED TO
%N(MNII); HOWEVER THEN, [MN/FE] LIMIT SHOULD BE EVEN HIGHER, THEREFORE MEANINGLESS.
1229$-$021 & 0.395 & 20.75{\scriptsize (0.07)}   &  $<$$-$1.32                 &$^{\rm 7}$& $>$$+$0.85                    &  ...                         &   $>$$+$0.49                     & $>$$+$0.56                   &  ...                             & n,o        \\%2.00,0.16
1247$+$267 & 1.223 & 19.87{\scriptsize (0.09)}   &  $-$1.41{\scriptsize (0.10)}& &  $+$0.39{\scriptsize (0.16)}  &  $+$0.18{\scriptsize (0.11)} &      $-$0.17{\scriptsize (0.07)} &  $+$0.30{\scriptsize (0.11)} &  ...                             & p          \\%0.10
1328$+$307 & 0.692 & 21.25{\scriptsize (0.06)}   &  $-$1.81{\scriptsize (0.12)}& &  $+$0.61{\scriptsize (0.14)}  &  $+$0.06{\scriptsize (0.14)} &   $<$$-$0.38                     &  ...                         &  ...                             & n,q,r      \\%1.30
1351$+$318 & 1.149 & 20.23{\scriptsize (0.10)}   &  $-$1.00{\scriptsize (0.11)}& &  $+$0.76{\scriptsize (0.11)}  &  $+$0.19{\scriptsize (0.11)} &      $-$0.04{\scriptsize (0.07)} &  $+$0.57{\scriptsize (0.11)} &  ...                             & p          \\%0.10
1354$+$258 & 1.420 & 21.54{\scriptsize (0.06)}   &  $-$2.02{\scriptsize (0.08)}& &  $+$0.43{\scriptsize (0.11)}  &  $+$0.23{\scriptsize (0.07)} &      $-$0.29{\scriptsize (0.07)} &  $+$0.30{\scriptsize (0.11)} &  ...                             & p          \\%0.10
1622$+$238 & 0.656 & 20.36{\scriptsize (0.10)}   &  $-$1.27{\scriptsize (0.16)}& &  ...                          &  ...                         &   $<$$-$0.20                     &   ...                        &      $+$0.41{\scriptsize (0.16)} & b,k,s      \\%1.70
1946$+$769 & 1.738 & ...$^{\rm ~1}$              &  $-$1.05                    & &  $<$$+$0.23                   &  $+$0.15{\scriptsize (0.07)} &      $-$0.36{\scriptsize (0.09)} &  $+$0.26{\scriptsize (0.05)} &  ...                             & f, t       \\
2128$-$123 & 0.430 & 19.37{\scriptsize (0.08)}   &  $>$$-$0.78                 & &  ...                          &  ...                         &   $<$$-$0.03                     &   ...                        &   $<$$-$0.29                     & $\star$,b  \\%0.16
2206$-$199 & 0.752 & ...$^{\rm ~1}$              &  $-$0.90                    & &  ...                          &  ...                         &      $-$0.10{\scriptsize (0.06)} &   ...                        &      $+$0.27{\scriptsize (0.04)} & u          \\%0.10
\hline
\end{tabular}
%}
\\
The numbers in parentheses are standard deviations.\\
$^{\rm 1}$ The total neutral hydrogen column density of these absorbers is
unknown, but since
$w_r($Fe\,{\sc ii}\,$\lambda$2600$)/w_r($Mg\,{\sc ii}\,$\lambda$2796$)\ga 0.6$,
they are at the very least moderate DLA systems. For these absorbers, a value
$\log N($H\,{\sc i}$)=20$ was adopted to estimate [Fe/H].\\
%Note that [Fe/H$]=-1.1$ is typical of this intermediate-redshift sample.\\
$^{\rm 2}$ All the detected components of Fe\,{\sc ii} were included to
estimate [Fe/H], whereas it is not the case for the abundance ratios (see
Sect.~\ref{elabs}). The difference between the values of [Fe/H] computed with
all the components of Fe\,{\sc ii} or with only those used to derive the
abundance ratios is important ($>0.12$ dex) only for the DLA
systems toward Q\,0551$-$366, Q\,1104$-$181 and Q\,1122$-$168.\\
$^{\rm 3}$ $z_{\rm abs}=1.9621$: [Fe/H$]=-1.26\pm 0.09$.\\
$^{\rm 4}$ $z_{\rm abs}=1.6615$: [Fe/H$]=-1.87\pm 0.01$.\\
$^{\rm 5}$ $z_{\rm abs}=0.6819$: [Fe/H$]=-1.83\pm 0.06$ (components $3+4$).\\
$^{\rm 6}$ $z_{\rm abs}=0.6826$: [Fe/H$]=-2.23\pm 0.05$ (component 9).\\
$^{\rm 7}$ The strong Fe\,{\sc ii} lines of this multiple absorption system
are saturated. A curve of growth analysis of
the Fe\,{\sc ii}\,$\lambda\lambda$2586,2600 lines
yields $\log N($H\,{\sc i}$)=14.47$, thus [Fe/H]=$-1.79$. \\
{\sc References:} ($\star$) This work;
(a) \citealt{2000ApJ...532...65P};
(b) \citealt{2000ApJS..130...91C};
(c) \citealt{1997A&A...327...47V};
(d) \citealt{1986MNRAS.220..833B};
(e) \citealt{1982MNRAS.200.1091B,1985ApJ...288..580B};
(f) \citealt{1996ApJS..107..475L};
(g) \citealt{1994A&A...283..759P};
(h) \citealt{1998A&A...337...51L};
(i) \citealt{2002A&Asubmitted..L};
(j) \citealt{1995ApJ...451L..13M};
(k) \citealt{2000ApJS..130....1R};
(l) \citealt{1999ApJ...513..598L};
(m) \citealt{2000A&A...363...69D};
(n) \citealt{1998A&A...333..841B};
(o) \citealt{1992ApJ...391...48L};
(p) \citealt{1999ApJ...510..576P};
(q) \citealt{1994ApJ...421..453C};
(r) \citealt{1992ApJ...399L.121M};
(s) \citealt{1997ApJ...480..568S};
(t) \citealt{1995ApJ...447..597L};
(u) \citealt{1997ApJ...474..140P}.
\label{ablitab}
\end{table*}

%------------------------------------------------------------------------------

\begin{table*}
\caption[]{Mean abundance ratios observed in DLA systems, along SMC lines
of sight, and in SMC and Galactic stars:}
\begin{tabular}{cccccccc}
\hline
Ratio & \multicolumn{2}{c}{DLA systems$^{\rm a,b}$}        & \multicolumn{2}{c}{SMC$^{\rm b,c}$} & \multicolumn{2}{c}{Galactic stars$^{\rm d}$}\\
      & $0.3<z_{\rm abs}\le 1.15$ & $1.15<z_{\rm abs}<2.2$ & ISM & stars                         & [Fe/H$]>-0.5$ & [Fe/H$]\sim -1$             \\
\hline
\ [Mn/Fe] &  $-$0.05$\pm$0.28 (14) & $-$0.18$\pm$0.22 (11) & $-$0.07$\pm$0.05
(3) & $+$0.04$\pm$0.17 & $-$0.06$\pm$0.05 & $-$0.26$\pm$0.05\\
\ [Si/Fe] &  $+$0.39$\pm$0.21 ( 5) & $+$0.36$\pm$0.13 ( 8) & $+$0.82$\pm$0.25
(3) & $+$0.14$\pm$0.11 & $+$0.08$\pm$0.05 & $+$0.28$\pm$0.05\\
\ [Ti/Fe] &  $+$0.12$\pm$0.27 ( 8) & $+$0.31$\pm$0.07 ( 3) & $+$0.28$\pm$0.37
(2) & $+$0.16$\pm$0.10 & $+$0.07$\pm$0.10 & $+$0.23$\pm$0.10\\
\ [Cr/Fe] &  $+$0.14$\pm$0.09 ( 6) & $+$0.15$\pm$0.06 ( 8) & $+$0.12$\pm$0.04
(3) & $+$0.10$\pm$0.12 & $+$0.00$\pm$0.10 & $-$0.02$\pm$0.05\\
\ [Zn/Fe] &  $+$0.48$\pm$0.26 ( 9) & $+$0.45$\pm$0.25 ( 7) & $+$0.73$\pm$0.20
(3) & $+$0.08\phantom{$\pm$0.00} & $+$0.05$\pm$0.10 & $+$0.12$\pm$0.10\\
\hline
\end{tabular}\\
$^{\rm a}$ Results from this work. Most detection limits are
significant enough (see Fig.~\ref{abobs}) to be included in the statistics as
detections.\\
%The abundance ratios observed in DLA systems at $z_{\rm abs}<1.7$ are
%very similar to those observed in DLA systems at higher redshifts;\\
$^{\rm b}$ The number of lines of sight involved in the computation is given
in parentheses.\\
$^{\rm c}$ Elemental ratios observed in the ISM of the SMC (lines of sight
toward Sk\,155, Sk\,78 and Sk\,108; \citealt{2001ApJ...554L..75W}) or in
SMC stars (appendix of \citealt{1997ApJ...489..672W}, and refs. therein).
Differences in the adopted Solar system abundances are accounted for.\\
$^{\rm d}$ Elemental ratios observed in Galactic halo ([Fe/H$]\sim -1$) and
Galactic thin disk ([Fe/H$]>-0.5$) stars \citep[e.g.][ and refs.
therein]{2000AJ....120.2513P}.\\
\label{arcetab}
\end{table*}

%------------------------------------------------------------------------------

\clearpage

%------------------------------------------------------------------------------
\end{document}